\documentclass[noshowpacs,amsmath,twocolumn,superscriptaddress,8pt,aps,prb]{revtex4-1} 
\usepackage{setspace}
\usepackage{amsmath}
\usepackage{breqn}
\usepackage{graphicx}
\usepackage[nearskip,margin = 0pt]{subfig}
\usepackage{verbatim} 
\usepackage{mathtools} 
\usepackage{bm} 
\usepackage{epstopdf}
\usepackage{xcolor}
\usepackage[normalem]{ulem} 
\usepackage{ragged2e}
\usepackage{array}
\usepackage[resetlabels,labeled]{multibib}
\DeclareGraphicsExtensions{.pdf,.eps,.png,.jpg,.mps}

\usepackage[colorlinks=true,allcolors=blue]{hyperref}

\DeclarePairedDelimiter{\ket}{\lvert}{\rangle}
\DeclarePairedDelimiter{\bra}{\langle}{\rvert}
\DeclarePairedDelimiter{\mean}{\langle}{\rangle}
\DeclarePairedDelimiter{\abs}{\lvert}{\rvert}

\DeclareMathOperator{\bigO}{\mathit O}
\DeclareMathOperator{\tr}{tr}

\newcommand{\vac}{\bm{0}}

\newcommand\mbsout{\bgroup\markoverwith{\textcolor{violet}{\rule[0.5ex]{2pt}{0.4pt}}}\ULon}

\renewcommand{\figurename}{Figure}

\begin{document}
\title{Mesoscopic cavity quantum electrodynamics with phase-disordered emitters \\in a Kerr nonlinear resonator}

\author{Daniil M. Lukin$^{*,1,7,\dagger}$, Bennet Windt$^{*2}$,
Miguel Bello$^{*2}$, Dominic Catanzaro$^{*1}$, Melissa A. Guidry$^{1}$,  Eran Lustig$^{1}$, Souvik Biswas$^{1}$, Giovanni Scuri$^{1}$, Trung Kien Le$^1$, Joshua Yang$^1$, Arina A. Nikitina$^3$, Misagh Ghezellou$^4$, Hiroshi Abe$^5$, Takeshi Ohshima$^{5,6}$, Jawad Ul-Hassan$^4$, and Jelena Vu\v{c}kovi\'{c}$^{1,\dagger}$\\
\vspace{+0.05 in}
$^1$E. L. Ginzton Laboratory, Stanford University, Stanford, CA 94305, USA.
\\
$^2$Max-Planck-Institute for Quantum Optics, Hans-Kopfermann-Str. 1, 85748 Garching, Germany
\\
$^3$Department of Molecular Cellular and Developmental Biology, University of California, Santa Barbara, California, USA.
\\
$^4$Department of Physics, Chemistry and Biology, Link\"oping University, SE-58183, Link\"oping, Sweden
\\
$^5$National Institutes for Quantum Science and Technology, Takasaki, Gunma 370- 1292, Japan
\\
$^6$Department of Materials Science, Tohoku University, Sendai 980-8579, Japan
\\
$^7$John A. Paulson School of Engineering and Applied Sciences, Harvard University, Cambridge, MA, USA
\\
$^*$These authors contributed equally 
\\
$^\dagger$dlukin{\makeatletter @\makeatother}seas.harvard.edu, jela{\makeatletter @\makeatother}stanford.edu
}

\begin{abstract}
The field of cavity quantum electrodynamics (QED) has seen a recent resurgence of interest in few- and many-body physics owing to the realization that the breaking of symmetries and the presence of disorder can give rise to entirely new phenomena \cite{lei2023many,yan2023superradiant}. Here we demonstrate a few-emitter cavity QED system capable of realizing new Hamiltonians in quantum optics based on breaking of symmetries and the realization of an \textit{in situ} Kerr nonlinearity. Our experiment relies on a high-finesse silicon carbide whispering gallery mode resonator hosting an ensemble of silicon vacancy color centers \cite{lukin2023two}. The simultaneous presence of spectral and spatial disorder of the mesoscopic atom system gives rise to emergent chirality, and the  optical nonlinearity of the silicon carbide host crystal enables the observation of atom-photon correlations induced by a four-photon nonlinear process. This work demonstrates the potential for solid state defect systems to realize emerging proposals \cite{leroux2018enhancing,groszkowski2020heisenberg, lau2025efficient} and to study fundamental physics in quantum electrodynamics.
\end{abstract}

\maketitle

\section*{Introduction}
In recent years, the study of quantum many-body physics has been advanced by experimental achievements in analog quantum simulation using synthetic quantum matter, such as trapped ions~\cite{monroe_programmable_2021, blatt_quantum_2012}, superconducting qubits~\cite{Zhang2023,guo_observation_2021}, (ultra-)cold atoms in optical lattices~\cite{gross_quantum_2017,bloch_many-body_2008,bloch_quantum_2012}, or Rydberg tweezer arrays~\cite{browaeys_many-body_2020,de_leseleuc_observation_2019,semeghini_probing_2021,bernien_probing_2017,ebadi_quantum_2021}. An ever-improving level of tunability across these platforms has opened the door to the exploration of controllable complexity in many-body systems, laying the groundwork for a deeper understanding of emergent phenomena. 

Within the purview of quantum optics, it is desirable to develop similarly controllable experimental platforms to study driven-dissipative many-body systems out of equilibrium. Especially in the presence of long-range photon-mediated interactions, such systems display rich physics, such as collective dissipative dynamics~\cite{wang2020,yan2023superradiant}, non-equilibrium phase transitions~\cite{kroeze_replica_2023,wang1973},
and emergent exotic spin models~\cite{norcia2018,davis2019}. Long-range interactions are naturally realized in the setting of cavity QED (CQED), which has recently garnered renewed interest in settings where the cooperative light-matter coupling competes with disorder, leading to the emergence of entirely new phenomena. Such phenomena have already been observed in single-mode CQED for spectral~\cite{lei2023many} and spatial~\cite{yan2023superradiant} disorder. Furthermore, the presence of even a single additional optical mode can significantly alter the properties of these light-matter interfaces~\cite{bechler2018passive, lukin2023two,vaidya2018tunable}.

To date, experimental realizations of CQED systems have predominantly focused either on large atomic ensembles ($N\sim 10^3-10^5$) amenable to a semi-classical description~\cite{kroeze_replica_2023,braverman_near-unitary_2019,hosten_quantum_2016,schleier-smith_squeezing_2010,lei2023many}, or very few individual atoms ($N\sim2$)~\cite{evans_photon-mediated_2018, lukin2023two}. Recently, deterministic loading of up to 10 atoms into a cavity has been demonstrated\cite{liu2023realization}, albeit in a highly symmetric configuration. Meanwhile, the regime of intermediate $N$ with individual atom control and tailored asymmetric tunable interactions remains largely unexplored experimentally, but provides a considerable theoretical challenge, and may prove crucial for developing a “bottom-up” understanding of strongly-correlated many-body quantum optical phenomena.

Here, we report on the study of this mesoscopic regime in a two-mode CQED system based on artificial atoms in silicon carbide (SiC)~\cite{lukin20204h, lukin2023two}. Through advances in nanofabrication, we reach the strong-coupling regime between a single atom and a cavity, and these strong interactions enable us to observe multi-emitter interference effects of $N\sim10$ atoms. The geometry of the cavity allows for independent spatial addressing of individual atoms and the extraction of Hamiltonian parameters, crucial for future atom-by-atom control in such systems. The intrinsic stability of phase and coupling strength in the solid state enable the use of this system to study phase and spectral disorder and their role in emergent phenomena, specifically in the suppression of photon correlations and the emergence of steady-state chirality. Finally, we leverage the unique advantage of the solid-state resonator --- its strong material Kerr nonlinearity \cite{guidry2020optical} ---  to realize a parametric drive term, where the atom ensemble is driven directly by the spontaneous photon pair generation in the resonator, opening the door to the observation of numerous effects based on emitter interactions with optical nonlinear processes \cite{carmichael1987resonance, parkins1993spectral, leroux2018enhancing,groszkowski2020heisenberg, le2024cavity}.

\section*{The device}

\begin{figure*}[t]
\includegraphics[width=\textwidth]{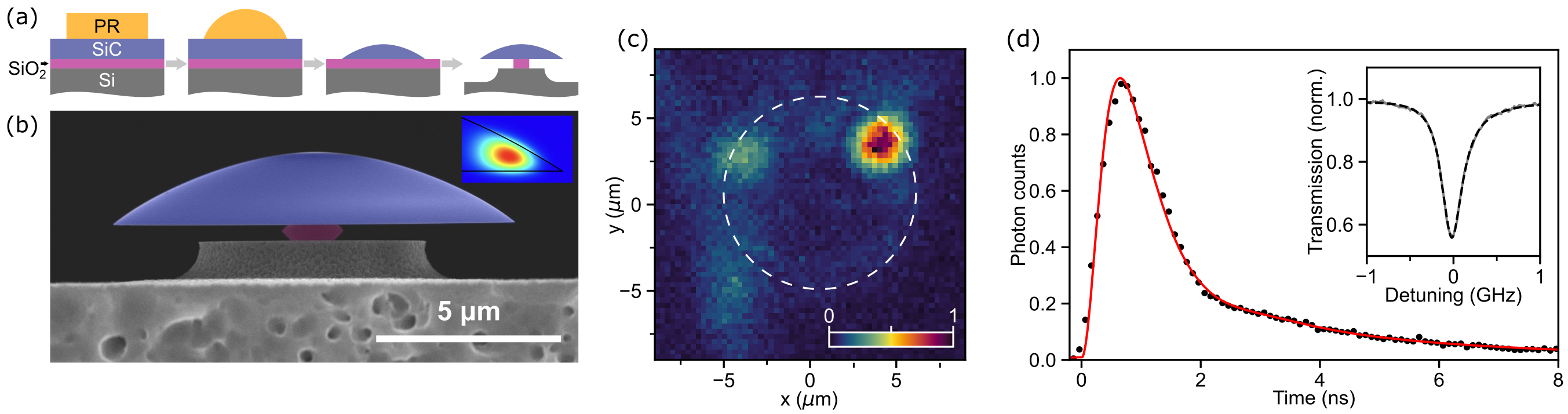}
\centering
\caption{\textbf{Device for strong coupling of multiple color centers to a whispering-gallery optical mode.} \textbf{(a)} Fabrication flow for high-finesse WGM resonators: Photoresist (PR) is patterned on silicon carbide on insulator; reflow is performed; PR pattern is transferred into SiC; device is undercut. \textbf{(b)} Scanning electron micrograph of the resonator profile. Inset: The profile of electric field intensity for the fundamental quasi-TM mode. \textbf{(c)} Raster scan via an excitation laser showing cavity on-resonance with a single strongly-coupled artificial atom. \textbf{(d)} Temporal evolution dynamics starting with the atom in the excited state, corresponding to $\{g,\gamma,\kappa\} = \{202, 65, 289\} \text{MHz}$ in presence of spectral diffusion. Inset shows the resonator mode, with intrinsic Q of $1.3\cdot^6$.}
\label{fig_single_emitter}
\end{figure*}

In order to reach the strong coupling regime with single artificial atoms, we develop a direct pattern transfer process from photoresist to SiC that enables small-diameter, low-roughness whispering gallery mode (WGM) resonators. A WGM resonator is typically considered sub-optimal for realizing strong light-matter interactions\cite{lukin2023two} due to its large mode volume ($V$) compared to the subwavelength confinement attainable in photonic crystals\cite{evans_photon-mediated_2018}. In terms of raw atom-cavity coupling rate $g \propto 1/\sqrt V$, this is indeed true. However, in the strong atom-cavity coupling regime, the magnitude of $g$ must be considered relative to both the atom and cavity decoherence rates ($\gamma$ and $\kappa$, respectively). In this context, WGM resonators, unmatched in attainable photon storage times, are quite appealing, since typically $\kappa$ is the limiting factor to attaining strong coupling. In our fabricated devices, a low surface roughness of 3 \r{A} RMS with a radius of curvature as small as 8.3~\textmu m is achieved, which results in a steep wedge angle that minimizes photon scattering (reduce $\kappa$) and reduces interactions of the atoms with surface noise. The fabrication flow and scanning electron image of a completed device are shown in Fig.~\ref{fig_single_emitter}(a),(b) (see Methods for details). To couple to the resonator, a traditional silica fiber-taper interface is ineffective due to the large index mismatch between SiC and SiO$_2$ and the mechanical instability in presence of vibrations of the closed-cycle cryostat. To address these challenges, we utilize a cryo-compatible and mechanically-compliant photonic microprobe. The microprobe consists of a single-mode SiC waveguide (fabricated with electron-beam lithography) interfaced directly to tapered single-mode fibers. For maximum mechanical stability, the waveguide is kept in contact with the resonator during measurement. Modifying the contact point between the waveguide and the resonator allows for wide and precise control of the waveguide coupling rate $\kappa_C$ as well as the coupling ideality\cite{pfeiffer2017coupling}. Details of the waveguide probe device are described in Ref. \cite{catanzaro2023cryogenic}.

The artificial atoms used in this work are the silicon vacancy (V\textsubscript{Si}) color centers\cite{liu2024silicon}. A monoatomic defect in a uniaxially symmetric 4H-SiC crystal, the V\textsubscript{Si} exists in only two lattice configurations, which are spectrally distinct. We focus only on the cubic configuration (\textit{k}-V\textsubscript{Si}), which has an optical transition at 916~nm, and thus treat all emitters as identical dipoles with an inhomogeneous spectral detuning. Their out-of-plane dipole moment couples optimally to the TM modes of the resonator. To generate the artificial atoms in the resonator, the crystal is uniformly irradiated with high-energy electrons prior to device fabrication (see Methods for details).

To map out the cavity-coupled emitters, a free-space excitation beam is rastered over the resonator while detecting emission into the waveguide. The resulting intensity map reflects both the location and the strength of atom-cavity coupling as well as the spectral overlap with the cavity mode. By tuning the cavity at a slow constant rate, a map of spatial and spectral distribution is obtained (Supplementary Video 1). 

Prior to studying multi-emitter effects, we consider the single-emitter CQED system, by tuning the resonator away from center of the atom's frequency distribution, where only one atom is resonant with the cavity mode. Figure~\ref{fig_single_emitter}(c) shows a scan on a device where only one emitter is dominantly coupled to the cavity mode. An undercoupled transmission scan of the cavity mode (inset of Fig.~\ref{fig_single_emitter}(d)) reveals a loaded (intrinsic) quality factor $Q_l = 1.13\cdot10^6$ ($Q_i = 1.29\cdot10^6$). To observe the temporal dynamics in the system, we excite the emitter from above with an above-resonant (wavelength 780 nm) picosecond mode-locked laser while monitoring the emission into the waveguide on a single-photon detector. The resulting dynamics, shown in Fig.~\ref{fig_single_emitter}(d), correspond to the system evolving with initial conditions of cavity and emitter in the ground and excited state, respectively. Since the emitter undergoes spectral diffusion in presence of above-resonant light, the observed dynamics are an average of many instances of the system with varying emitter-cavity detuning (see Methods). While this averaging suppresses the direct observation of Rabi oscillations, the fit reveals an emitter-cavity coupling rate $g = 202$~MHz. The ratio of these rates signifies the strong coupling regime, $4g/\kappa_I =  3.2 > 1$, the largest reported for an atomic defect coupled to a cavity \cite{bhaskar2020experimental}. We perform a Hong-Ou-Mandel (HOM) measurement of photon indistinguishability. From a raw visibility of 0.76(4), we obtain an upper bound on the magnitude of dephasing ($\gamma' \leq 39$~MHz) and verify the absence of spectral diffusion on the timescale of the inter-pulse delay (Extended Data Fig.~\ref{fig_indistinguishability}).

\begin{figure*}[t]
\includegraphics[width=\textwidth]{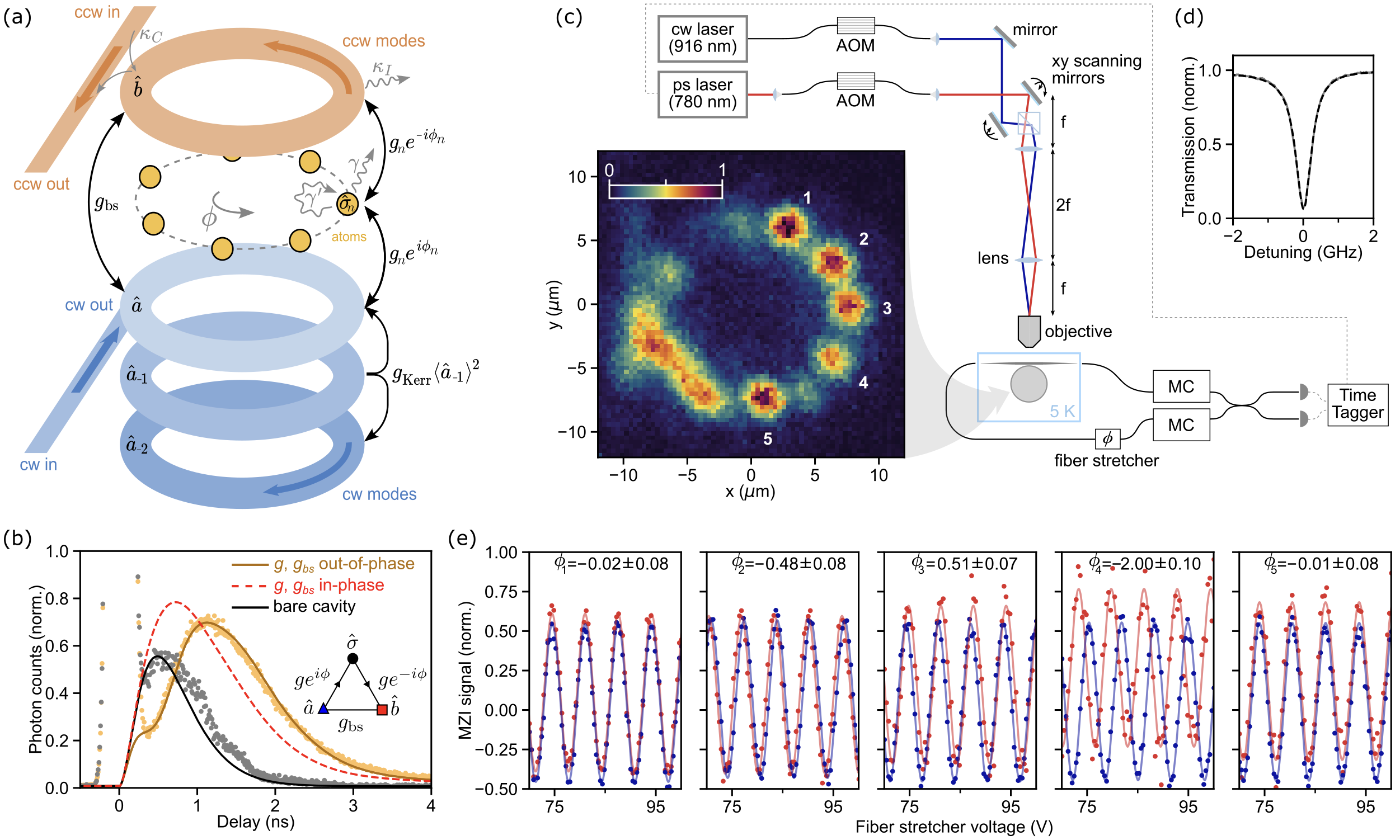}
\centering
\caption{\textbf{Model Hamiltonian and phase relations of atom-cavity and cavity-cavity coupling.} \textbf{(a)} The ensemble of emitters $\hat\sigma_j$ couples to the clockwise ($\hat a$) and counterclockwise ($\hat b$) cavity modes with phase $\phi_j$ and coupling strength $g_j$ which depend on the position of the emitter in the resonator. Modes $\hat a, \hat b$ may also couple directly via a backscattering rate $g_{\text{bs}}$. The Kerr nonlinear coupling between three clockwise modes $\hat a$, $\hat a_{-1}$, and $\hat a_{-2}$ is proportional to the single-photon nonlinear shift $g_{\text{Kerr}}$. The Markovian bath coupling is indicated in grey. All cavity modes share the same waveguide and intrinsic decay rates ($\kappa_C$ and $\kappa_I$). The emitters' spontaneous emission and dephasing rates are denoted by $\gamma$ and  $\gamma'$, respectively. \textbf{(b)} Single-emitter back-scattering measurement revealing the effect of atom phase relative to $g_{\text{bs}}$ via destructive interference of excitation transfer from $\hat a$ to $\hat b$. Inset: Schematic representation of the system Hamiltonian. \textbf{(c)} Experimental setup for measurement of individual atoms' phases $\phi_j$ in the cavity. Inset shows the distribution of atoms in the cavity for one cavity detuning \textbf{(d)} Cavity transmission scan (detuning relative to  327,108.8~GHz). \textbf{(e)} Interferometric measurement of the phases of the five spots labeled in (c). Atom fluorescence and reference signal are shown in red and blue, respectively, and the atom-cavity phase $\phi_i$ is indicated in each plot (in radians).}
\label{fig_hamiltonian}
\end{figure*}

\section*{The Model}
Having benchmarked the performance of the single-emitter CQED system, we proceed to the full multi-emitter system. The Hamiltonian, visualized in Fig.~\ref{fig_hamiltonian}(a), can be decomposed as $\hat{H}=\hat{H}_e+\hat{H}_{\rm cav}+ \hat{H}_{\rm int}$, with ($\hbar=1$)
\begin{gather*}
    \hat H_e = \sum_{n=1}^N \omega_n^e \hat\sigma^\dag_n \hat\sigma_n \\
    \hat H_{\rm cav} = \omega_{\rm cav}\left(\hat a^\dag \hat a + \hat b^\dag \hat b\right) + g_{\rm bs}\left(\hat a^\dag \hat b + {\rm H.c.}\right) \\
    \hat H_{\rm int} = \sum_{n=1}^N \frac{g_n}{\sqrt{2}} \left(e^{i\phi_n}\hat\sigma^\dag_n\hat a + e^{-i\phi_n}\hat\sigma^\dag_n\hat b\right) + {\rm H.c.}
\end{gather*}
The terms $\hat H_e$ and $\hat H_{\rm cav}$ capture the bare emitter and cavity degrees of freedom, respectively. We denote the lowering operator for the $n$-th emitter by $\sigma_n = \ket{g}\!\bra{e}_n$, and its optical transition frequency by $\omega_n^e$; $\hat a$ and $\hat b$ denote the annihilation operators of the clockwise (cw) and counter-clockwise (ccw) modes that couple to the emitter ensemble, respectively, which are degenerate, with frequency $\omega_{\rm cav}$. Note that $\hat H_{\rm cav}$ contains an additional direct coupling at strength $g_{\rm bs}$ between the modes (backscattering), which arises from the breaking of rotational symmetry due to geometric imperfections of the resonator and proximity of the waveguide probe. The term $\hat H_{\rm int}$ represents the emitter-cavity interaction, with the $n$-th emitter coupling to each of the two cavity modes with equal, real-valued strength $g_n$ and a position-dependent coupling phase $\phi_n$. The full model $\hat H$ therefore resembles a Tavis-Cummings (TC) model, with two cavity modes instead of one, and with complex coupling strengths.

In addition, the WGM system features a four-photon nonlinear (Kerr) coupling between cw modes of the form $\hat a_m \hat a_m \hat a^\dagger_{m-1}  \hat a^\dagger_{m+1}$. In this work, we implement a non-degenerate pair generation process driven by a strong coherent pump in the mean-field undepleted-pump regime, where modes $\hat a, \hat a_{-2}$ are coupled via a pulsed parametric drive with amplitude $\Omega_{\text{Kerr}}(t) = g_{\text{Kerr}}\langle \hat a_{-1} \rangle ^2$. This is represented by the additional Hamiltonian term
\begin{equation*}
    \hat H_{\rm Kerr} = \omega_{-2}\hat a^\dag_{-2}\hat a_{-2}+g_{\rm Kerr} \left(\mean{\hat a_{-1}}^2 \hat a^\dag \hat a^\dag_{-2} + \mathrm{H.c.}\right)\,,
\end{equation*}
where $\omega_{-2}$ denotes the frequency of mode $\hat{a}_{-2}$.

We also account for the interaction of the system with the continuum of free-space modes. The intrinsic loss rate of the cavity modes is denoted by $\kappa_I$, while the emitter spontaneous emission and dephasing rates are $\gamma_n$ and $\gamma'_n$, respectively. The cavity modes couple to cw- and ccw-propagating waveguide modes with rate $\kappa_C$. Due to the high finesse of the cavity, modes $\hat a, \hat a_{-1}, \hat a_{-2}$ decay into independent Markovian baths. Furthermore, the atoms are amenable to individually-controllable incoherent (above-resonant) excitation at a rate $\gamma^{\rm ex}_n$. The full Liouvillian describing the system evolution thus reads
\begin{equation*}
    \mathcal{L}\hat{\rho} = -i[\hat H, \hat\rho] + \mathcal{D}_e \hat\rho + \mathcal{D}_{\rm cav} \hat\rho \,,
\end{equation*}
with dissipative terms given by 
\begin{gather*}
    \mathcal{D}_e = \sum_{n} \left(\gamma_n \mathcal{D}[\hat\sigma_n] + \gamma'_n \mathcal{D}[\hat\sigma_n^\dag \hat\sigma_n] + \gamma^{\rm ex}_n\mathcal{D}[\hat\sigma^\dag_n]\right) \\
    \mathcal{D}_{\rm cav} = \kappa \left(\mathcal{D}[\hat a] + \mathcal{D}[\hat b] + \mathcal{D}[\hat a_{-1}] + \mathcal{D}[\hat a_{-2}]\right) \,,
\end{gather*}
where $\mathcal{D}[\hat x]\hat{\rho}=\hat x\hat{\rho}\hat x^\dagger-\{\hat x^\dagger \hat x,\hat{\rho}\}/2$ is the Lindblad dissipation superoperator associated to the jump operator $\hat x$ and $\kappa = \kappa_I + \kappa_C$.

The complex coupling phases $\phi_n$ in the interaction Hamiltonian $\hat{H}_{\rm int}$ play a central role in the photon transport through the system. At the level of a single atom, we can observe an interference effect akin to the Aharonov-Bohm effect whereby single photons are transferred from $\hat{a}$ to $\hat{b}$ either (i) directly via $g_{\rm bs}$ or (ii) by a cascaded process via the emitter, thereby acquiring a phase $e^{2i\phi}$. This is consistent with simulations, where we find that for $\phi = (0 \mod \pi)$ the interference is constructive (red dashed line in Fig.~\ref{fig_hamiltonian}(b)). To extract this phase in the single-emitter system, we coherently excite the device through the waveguide and measure the back-scattered photons. We observe instead a destructive interference and, using a three-parameter fit ($g$, $\phi$, and $g_{\text{bs}}$), we obtain $\phi = 0.35\pi$. If, on the other hand, $\sqrt{g_{\text{bs}}}$ greatly exceeds the collective ensemble-cavity coupling strength, it dominates the temporal back-scattering dynamics entirely (see Methods and Extended Data Fig.~\ref{fig_g_bs}). It is apparent that if $g_{\rm bs} = 0$, photon transport between modes $\hat a$ and $\hat b$ is mediated solely by the atoms. Another striking effect of the complex couplings is the emergence of \emph{chiral steady states}, which we discuss in detail in a later section.

In the multi-emitter system, this prompts the question of how the relative emitter coupling phases can be experimentally measured. While in principle the information about $\phi_n$ is present in the spatial map of the emitters (see Fig.~\ref{fig_hamiltonian}(c)), the high sensitivity of phase to position ($\sim1$~deg/nm) would require not only superresolution imaging but also precise knowledge of the resonator center point and the angular momentum mode number. Here we present a direct interferometric phase measurement, which requires no such prior knowledge. As illustrated in Fig.~\ref{fig_hamiltonian}(c), an above-resonant (780 nm) free-space laser excites individual fluorescent spots, and the cw and ccw emission directions are combined on a beamsplitter to form a fiber interferometer. In order to distinguish atom phase from the fluctuating phase of the fiber interferometer, a reference interference signal from a scattering point on the resonator is acquired simultaneously (stroboscopically) using a laser resonant with the cavity. This permits independent measurement of the phase (mod $\pi$) for each spatially-resolvable atom, as shown in Fig.~\ref{fig_hamiltonian}(e). 

These complex couplings to the two common cavity modes leads to photon-mediated interactions between the emitters.
In the special case of equal coupling phases $\phi_n = \phi$, for all $n$, and $g_{\rm bs} = 0$, re-expressing the emitter-cavity couplings in terms of standing modes $\{(e^{i\phi}\hat a + e^{-i\phi}\hat b)/\sqrt{2}, (e^{-i\phi}\hat a - e^{i\phi}\hat b)/\sqrt{2}\}$ de-couples one mode from the ensemble, constituting a simplification to a single-mode CQED system. Less trivial configurations give rise to a more complex interplay of the coupling phases in the cavity-mediated interactions, which are unique to the multi-mode setting. 
In the regime where $\kappa \gg g_n, \gamma_n, \gamma'_n, \gamma^{\rm ex}_n$ for $n=1,\dots,N$,
the cavity modes can be adiabatically eliminated~\cite{jaeger2022lindblad} to obtain an effective master equation 
\begin{equation*}
    \partial_t \hat\rho_e = -i[\hat H_{\rm eff}, \hat\rho_e] + \mathcal{D}_{\rm eff}\hat\rho_e    
\end{equation*} 
for the reduced emitter state $\hat \rho_e$, with (see Methods)
\begin{gather*}
    \hat H_{\rm eff} = \sum_{n} \Delta_n \hat\sigma^\dag_n \hat\sigma_n + \sum_{m,\,n} J_{mn} \hat\sigma^\dag_m \hat\sigma_n \,, \\
    \mathcal{D}_{\rm eff} \hat\rho_e = \mathcal{D}_e \hat\rho_e + \sum_{m,\,n} \Gamma_{mn} \left(\hat\sigma_m \hat\rho \hat\sigma^\dag_n - \frac{1}{2}\{\hat\sigma^\dag_n \hat\sigma_m, \hat \rho_e\}\right) \,,
\end{gather*}
and $\Delta_n = \omega_n^e - \omega_{\rm cav}$. In the case $g_{\rm bs} = 0$, the dissipative and coherent interaction strengths, $\Gamma_{mn}$ and $J_{mn}$, can be expressed, after a gauge transformation, as
\begin{equation*}
    \Gamma_{mn} = \sqrt{\Gamma_m \Gamma_n} \cos(\phi_m - \phi_n) \,, \ 
    J_{mn} = \frac{\Delta_m + \Delta_n}{2\kappa} \Gamma_{mn} \,.
\end{equation*}
Here, $\Gamma_n = \kappa g_n^2/(\Delta_n^2 + \kappa^2/4)$ corresponds to the Purcell enhancement of the $n$th-emitter decay rate. For $g_{\rm bs} \neq 0$, general formulae for $J_{mn}$ and $\Gamma_{mn}$ can also be obtained and display an intricate dependence on both the coupling phases $\phi_n$ and the detunings $\Delta_n$ (see Methods).

Master equations of the form above are ubiquitous in quantum optics. 
They describe the Markovian dynamics of quantum emitters in a wide variety of settings, including free space and photonic structures, wherein the properties of the electromagnetic environment lead to very different interaction matrices $J_{nm}$ and $\Gamma_{nm}$.
Interestingly, when the emitters are all identical except for their coupling phases, the collective decay rates in our model at $g_{\rm bs}=0$ closely resemble those of a set of identical emitters coupled to a linear waveguide~\cite{chang_cavity_2012,pichler_quantum_2015}, with the coupling phases playing a role analogous to the positions of the emitters along the waveguide. 
This makes our platform an interesting system to investigate phenomena typically associated with waveguide QED (wQED) settings, such as the recently proposed dynamical mirror (cw-ccw) symmetry breaking~\cite{cardenas2023many}. However, the coherent interactions in our systems are qualitatively different from those present in wQED setups, which invariably leads to further intriguing many-body phenomena beyond the wQED regime. Moreover, for $g_{\rm bs} \neq 0$, the coupling matrices $J_{nm}$ and $\Gamma_{nm}$ cannot be made real simultaneously through a gauge transformation, a situation thus far not explored much in literature.
Given the important role of such complex interactions in traditional condensed matter systems, giving rise for example to integer and fractional quantum Hall effect phases, one might expect a similarly important role in the physics of disordered many-body systems such as the one we study.

\begin{figure*}[t]
\includegraphics[width=1\textwidth]{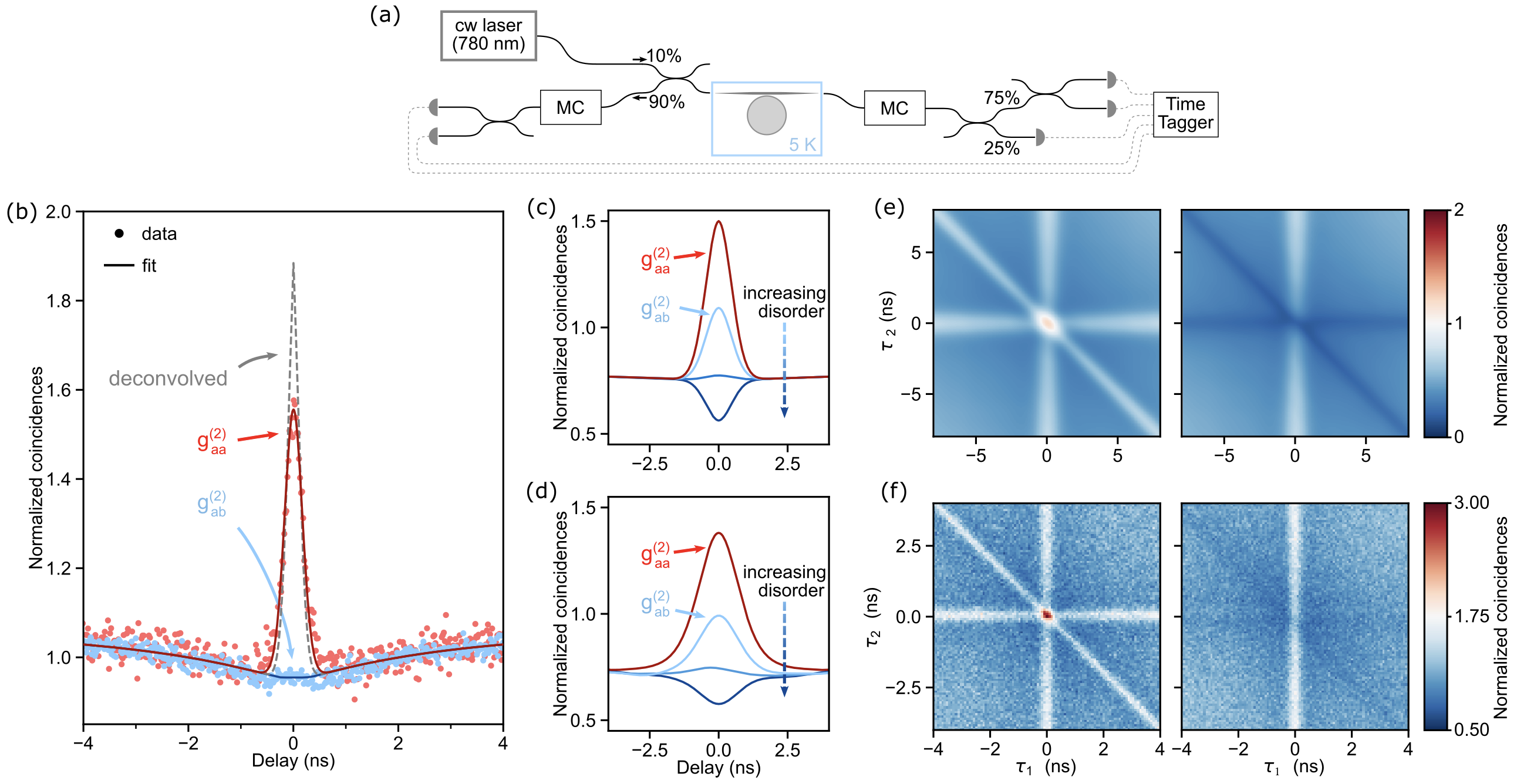}
\centering
\caption{\textbf{Steady-state photon correlations from a phase-disordered ensemble.} \textbf{(a)} Experimental setup for measurement of photon correlations. \textbf{(b)} Two-photon auto-correlation $g^{(2)}_{aa}(\tau)$ (cw,cw), and cross-correlation $g^{(2)}_{ab}(\tau)$ (cw, ccw). Fit to the bad-cavity model estimates  $N = 17.4$ coupled emitters with phase disorder parameter of $\xi_\phi = 0.07$. The correlated emission seen in auto-correlation is suppressed in cross-correlation due to phase disorder. \textbf{(c)} Theoretical photon correlations from $N=4$ uncorrelated identical emitters for different choices of coupling phases $\phi_n$, corresponding with $\xi_\phi\approx0.6, 0.3,0.06$. \textbf{(d)} Exact simulations for the same configurations as \textbf{(c)}. We observe clear qualitative agreement in the dependence on $\xi_\phi$. \textbf{(e)} Measured third-order correlations $g^{(3)}_{aaa}$ (left) and $g^{(3)}_{aab}$ (right). \textbf{(f)} Third-order correlations obtained from numerical simulations of the bad-cavity master equation, showing qualitative similarity in the role of $\xi_\phi<1$ on the bunching at zero time delay. In all simulations, we choose $\gamma_n=15~\mathrm{MHz}$, $\gamma_n'=40~\mathrm{MHz}$, $\gamma_n^{\rm ex}=0.3\gamma_n$, $g_n=150~\mathrm{MHz}$ and $\kappa=300~\mathrm{MHz}$, consistent with experimental parameter ranges.}
\label{fig_correlations}
\end{figure*}

\section*{Photon statistics from a phase-disordered few-emitter ensemble}

The many-body effects that arise from cavity-mediated emitter interactions can be observed in their steady-state photon correlations. For instance, second-order correlation functions can be used to probe the collective emission properties of the system~\cite{sipahigil_integrated_2016,machielse_quantum_2019,grim2019scalable,koong_coherence_2022,kim_super-radiant_2018}. We have previously demonstrated in the two-atom case~\cite{lukin2023two} that the relative coupling phases of the atoms dictate the nature of directional collective emission into the clockwise and counter-clockwise modes $\hat{a}$ and $\hat{b}$. The correlation measurements in our two-mode system thus display more complex features than those observed in single-mode settings.

Since $\kappa$ and $g$ in our system are comparable in magnitude, emitters that are simultaneously on-resonance with the cavity are expected to emit mutually-indistinguishable photons. To measure second- and third-order correlations ($ g^{(2)}(\tau),  g^{(3)}(\tau_1, \tau_2)$) in the system, we excite the resonator at a resonance around 780~nm with a continuous-wave laser, which drives the emitters incoherently. We detect cw and ccw emission via five superconducting nanowire single-photon detectors (SNSPDs), arranged as shown in Fig.~\ref{fig_correlations}(a). This configuration permits simultaneous observation of up to three emission events into a single direction as well as cross- and auto- two-photon correlations in both directions. We over-couple the cavity to the waveguide ($Q_L=3\cdot10^5$) to increase the photon collection efficiency. The measured two-photon auto- and cross-correlations are shown in Fig.~\ref{fig_correlations}(b). The auto-correlations $g_{aa}^{(2)}(\tau)$ display the well-known bunching behavior previously observed in the context of wQED with quantum dots~\cite{grim2019scalable,koong_coherence_2022,kim_super-radiant_2018} and V\textsubscript{Si} color centers~\cite{sipahigil_integrated_2016,machielse_quantum_2019}, and most recently in CQED with an ensemble of nitrogen vacancy centers in a fiber Fabry-P\'erot cavity~\cite{pallmann2023cavity}. The cross-correlations, however, are anti-bunched, with $g_{ab}^{(2)}(0) < 1$. This difference between the auto- and the cross-correlation is due to the phase disorder of the emitters in the two mode-degenerate resonator.

In fact, the suppression of the cross-correlation due to phase disorder can already be observed at the level of uncorrelated identical emitters. Under the more stringent bad-cavity condition $\sqrt{\kappa} \gg \sqrt{\gamma_n}, \sqrt{\gamma'_n}, \sqrt{\gamma^{\rm ex}_n}$, we find that for uncorrelated emitters, $g_{aa}^{(2)}(0)=2(1-1/N)$ while $g_{ab}^{(2)}(0)=1+\xi_\phi-2/N$ (see Methods), where the parameter $\xi_\phi = \sum_{ij} \cos(2(\phi_i - \phi_j))/N^2$ quantifies the phase disorder. Specifically, in the absence of varying coupling phases ($\xi_\phi=1$), bunching is not suppressed. In this case, even at non-zero time delays $g^{(2)}_{ab}(\tau) = g^{(2)}_{aa}(\tau)$. In the opposite limit of maximum disorder ($\xi_\phi = 0$), the cross-correlations are maximally suppressed. The theoretical dependence of the photon correlations from independent identical emitters on $\xi_\phi$ is illustrated in Fig.~\ref{fig_correlations}(c), and we show in Fig.~\ref{fig_correlations}(d) that this feature persists also in the full model. Indeed, this qualitative agreement reflects the challenge of discerning measurement-induced interference effects from truly cooperative emission exclusively from photon statistics~\cite{cygorek2023signatures}.

To model the experimental data, we derive an expression for correlations from independent emitters in the presence of a dark metastable state \cite{liu2024silicon} to account for the characteristic bunching at small time delays. We obtain best fit parameters $N = 17.4$ and $\xi_\phi = 0.07$, indicating a high degree of disorder expected from emitters with a random azimuthal distribution. 
However, we note that caution should be taken in interpreting $N$ as the experimentally-observed number of cavity-coupled emitters, due to the simplifying assumptions made in deriving the fit function, which we discuss in detail in the Methods. Firstly, we assume identical emitters subject to an identical spectral diffusion, for which we account only in an approximative way. Secondly, both the bad-cavity limit and the further assumption of uncorrelated emitters constitute significant approximations, departures from which would be expected, at the very least, at the short-time delay obscured by the detector jitter. While the low ensemble optical coherence of our system prevents us from observing definitive signatures of inter-emitter correlations, it is likely that correlations are present in the system, as suggested by the quantitative differences between the predictions of the full model (with comparable parameters to our experiment) and the model of uncorrelated emitters seen in Figs.~\ref{fig_correlations}(c),(d), respectively. Further experimental advances, most crucially the spectral stabilization\cite{anderson2019electrical} or post-selection\cite{van2024check} of emitters, will be necessary to observe definitively correlated multi-emitter states.

It is worth noting that the effect of phase disorder manifests also in higher-order photon correlations. In Figs.~\ref{fig_correlations}(e) and~\ref{fig_correlations}(f), we demonstrate, for instance, that the third-order auto- and cross-correlations likewise display bunching and disorder-induced anti-bunching behavior, respectively. Experimental access to these higher-order correlation functions could potentially prove useful in the future for reliably characterizing the structure of inter-emitter correlations or the lack thereof, and to observe multi-photon effects such as spontaneous symmetry breaking \cite{cardenas2023many}.

\begin{figure*}[t]
\includegraphics[width=\textwidth]{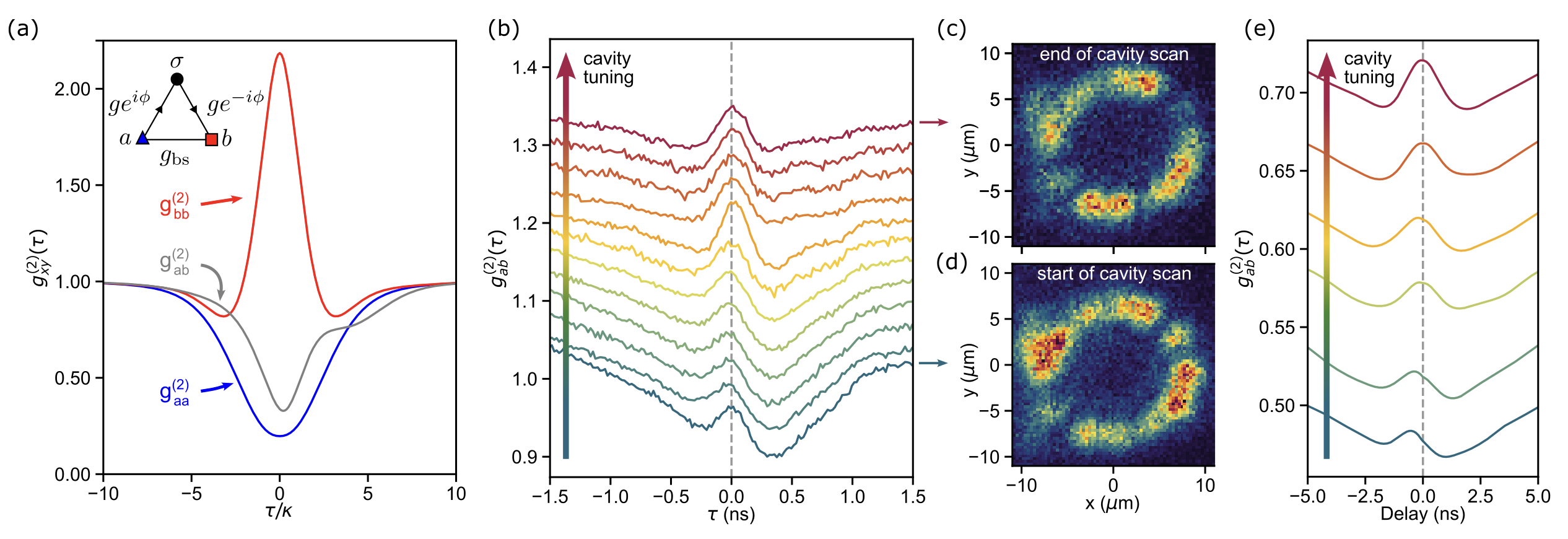}
\centering
\caption{\textbf{Emergence of steady-state chirality in the presence of spectral and phase disorder.} \textbf{(a)} The simplest example of chiral emission from a strongly-coupled atom-cavity system with inter-mode coupling. Chirality manifests as non-equivalence of the cw and ccw auto-correlations (red and blue) and as asymmetry of cross-correlation (grey). Simulation parameters are chosen as $\Delta = 0, \phi= \pi/4$, $g = 0.3\kappa$, $g_{\text{bs}} = 0.5\kappa$, $\gamma = 0.2\kappa$, and  $\gamma' = \gamma^{\rm ex} = 0.1\kappa$.  \textbf{(b)} We experimentally observe the transition between chiral and achiral emission in a multi-emitter ensemble via continuous tuning of the cavity frequency. Each trace is offset vertically by 0.03 for clarity. \textbf{(c,d)} Above-resonant fluorescence scan of the cavity at the end and the beginning of cavity scan, respectively. Reduction (increase) in an emitter intensity between start and end of scan indicates the cavity tuned closer (further) from the emitter center frequency. \textbf{(e)} Theoretical model with four emitters illustrating chirality behavior qualitatively similar to the experimental observation. Each trace is offset vertically by 0.04 for clarity.}
\label{fig_chirality}
\end{figure*}

\section*{Emergence of steady-state chirality in the presence of spectral and phase disorder}

Beyond the disorder-induced suppression of photon correlations, the presence of complex emitter-cavity couplings also results generally in a \emph{chiral steady state},\cite{lodahl2017chiral}
where the auto-correlation functions of photons emitted differ depending on the direction (cw or ccw), and where the cross-correlation function of photons emitted in opposite directions is asymmetric in the delay time.
This can be observed already in the simplest scenario, where a single emitter is coupled to the cavity modes, provided $g_{\rm bs} \neq 0$ (see Fig.~\ref{fig_chirality}(a)).
In that case, the Hamiltonian of the system is not invariant under the exchange of the cavity modes $\hat a \leftrightarrow \hat b$, which is equivalent to a change in the coupling phase $\phi \to -\phi$.
For $g_{\rm bs} = 0$, by contrast, the complex coupling phase can be removed by a gauge transformation (redefining $e^{i\phi} \hat a \to \hat a$, and $e^{-i\phi} \hat b \to \hat b$), so the steady state is not chiral in that case.
We also note that for a single emitter, the second-order correlation functions that can be computed from the emitter dynamics according to the bad-cavity master equation would all be the same. Therefore, in this case, any observed steady-state chirality indicates non-Markovian dynamics of the emitters.

On the other hand, for a system with more than one emitter, steady-state chirality can be seen to emerge quite generally within a Markovian description already at the level of uncorrelated emitters, arising from an interplay between spectral and phase disorder. Beyond the assumption of independent emitters, however, it is difficult in general to predict whether the steady state will be chiral or not (see Methods for details).
For example, the case of two emitters will not display chirality provided $g_{\rm bs} = 0$ and that the emitters are indistinguishable (aside from having different coupling phases), since a change in the sign of the phases can then be compensated by a permutation of the emitters.
For more than two emitters, the system does not have such a (weak) symmetry, except in finely tuned situations, and we expect a chiral steady state, even for indistinguishable emitters.
Interestingly, the presence of chirality appears to be intimately linked to the non-linearity of the emitters, as there are cases in which a homologous system where the emitters are replaced by bosonic modes will display symmetric correlations in the steady state whereas the original emitter system will not.

Experimentally, we find evidence of chirality by observing asymmetric cross-correlations $g_{\text{ab}}(\tau)$. Since the degree of chirality depends on the configuration of emitter detunings with respect to the cavity frequency, by tuning just a single parameter, namely the cavity detuning, one may observe the transition between chiral and achiral steady states. This is shown in Fig.~\ref{fig_chirality}(b): as the cavity is red-tuned, chirality disappears. By examining the above-resonant flourescence map of the emitters, one can see that the relative intensities of the emitters change significantly between the start and end of cavity tuning, indicative of variation of spectral detuning of the emitters (see Figs.~\ref{fig_chirality}(c),(d)). In other words, the variation of the cavity detuning essentially results in the enhanced participation of a subset of emitters in the emission process, whose spectral and phase relations determine the chirality. We are able to qualitatively reproduce this phenomenon for uncorrelated emitters (see Fig.~\ref{fig_chirality}(e)).

\begin{figure*}[t]
\includegraphics[width=\textwidth]{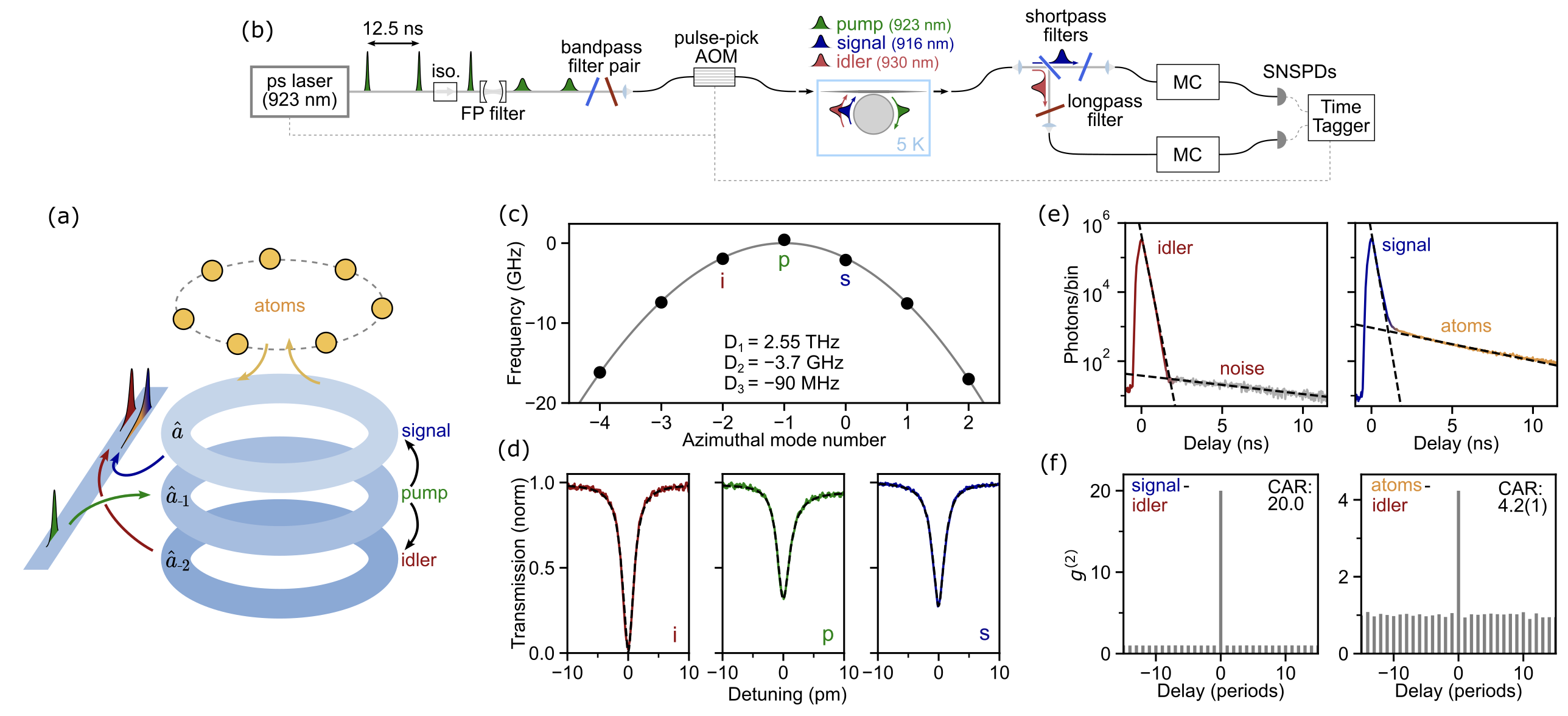}
\centering
\caption{\textbf{Atom-photon correlations via parametric drive of a Kerr resonator.} 
\textbf{(a)} Illustration of the system dynamics. The pump mode is driven to generate entangled photon pairs into the signal and idler modes. Whereas the idler photon is emitted from the cavity directly, the signal photon can be absorbed and then re-emitted by the emitter ensemble. \textbf{(b)} Diagram of the experimental setup. 
\textbf{(c)} The measured dispersion of the fundamental TM mode family of the resonator.  \textbf{(d)} Loaded Q factors of pump, signal, and idler of $3.7\cdot 10^5$, $3.9\cdot 10^5$, and $4.3\cdot 10^5$, respectively. Pump and signal modes are over-coupled to the waveguide.
\textbf{(e)} Temporal emission dynamics into the idler (left) and signal (right) modes. Dashed lines are exponential fits. Idler mode emission is characterized by a single dominant decay rate (180~ps), whereas the signal emission features two rates, fast (160~ps) and slow (4.6~ns). The latter corresponds to photons scattering from the atom ensemble. \textbf{(f)} Second-order photon correlation measurements for signal-idler and atoms-idler photon coincidences.}
\label{fig_nlo}
\end{figure*}

\section*{Parametric Kerr drive in a cavity QED system}

We now turn to the study of the interaction between the resonator Kerr nonlinearity and the artificial atoms. The high  Q and the strong optical nonlinearity of SiC render our system naturally suitable for this experiment. To observe the in-situ interaction of a parametric nonlinearity and a CQED system, we consider the simplest dynamical process, illustrated in Fig.~\ref{fig_nlo}(a). A pulsed coherent pump is injected into $\hat a_{-1}$, which drives signal-idler pair generation into modes $\hat a$ and $ \hat a_{-2}$. In absence of the atomic ensemble, the signal-idler pairs will decay from the resonator on the time-scale of the cavity lifetime. If, however, the atoms are coupled to $\hat a$, they may absorb the signal photon, and then re-emit it; this process takes place on the time-scale of the (cavity-enhanced) atomic decay rate. The system will thus exhibit correlations between idler photons and the emission from the atom ensemble.

For this experiment, we use the same resonator and transverse mode as in the multi-emitter experiments above. The experimental diagram is  illustrated in Fig.~\ref{fig_nlo}(b). The optical parametric process is driven by a pulse-shaped mode-locked laser. A 2~ps pulse train with repetition rate of 80~MHz is transformed into a 50~ps pulse train with repetition rate of 0.66~MHz using a Fabry-P\'erot (FP) cavity and an AOM. This accomplishes bandwidth-matching of the pump to the cavity mode and reduces system heating via the lower repetition rate. After passing through the device, the pump is spectrally filtered and the signal and idler are separated via a dichroic filter and sent to single-photon detectors. Photon arrival times are referenced to the pulsed laser clock.

The parametric pair generation process is realized between a pair of modes in the quasi-TM mode family of the resonators. The complementary modes in the family were identified via a high-precision laser transmission scan (Extended Data Fig.~\ref{fig_mzi}). The measured normal dispersion of $-3.7$~GHz, shown in Fig.~\ref{fig_nlo}(c), is consistent with finite-element model prediction. The strong normal dispersion of the resonator is not optimal for realizing an efficient FWM process due to substantial frequency mismatch. In order to reduce the detuning of the nonlinear process relative to $\kappa$, we increase the loaded Q-factor of the modes by over-coupling the resonator to the waveguide; this also increases the photon extraction efficiency. The pump, signal and idler mode transmission spectra are shown in Fig.~\ref{fig_nlo}(d). The pump mode is chosen to be a longer-wavelength mode relative to the atom-coupled mode to avoid above-resonant excitation of the V\textsubscript{Si} by the high-power laser. At a pump pulse energy of 15~pJ, accounting for detection and resonator out-coupling efficiency, we estimate that signal-idler photon pairs are generated at a rate of 0.01 pairs/pulse. Examining the temporal dynamics of photon emission from the resonator into the signal and idler modes, we observe that whereas the idler mode shows decay only at the fast timescale of the cavity lifetime (180 ps), the signal mode features an additional slow decay corresponding to the Purcell-enhanced lifetime of the emitter ensemble (4.6~ns). While this immediately suggests the observation of signal photons scattered by the atomic ensemble, caution must be used in attributing the atomic emission to the parametric drive: At the high pump power levels, other pathways, such as two-photon excitation, may have contribution to the atomic fluorescence. A proof that the atom ensemble is driven via parametric pair generation can be obtained by examining the photon correlations between the atomic emission and the idler photons. In Fig.~\ref{fig_nlo}(f), we compare the second-order correlation $g^{(2)}$ between the idler mode and the slow and fast emission into the signal mode. The fast emission (within $<0.8$~ns of the pump pulse) corresponds to direct decay of a signal photon from the cavity. The slow emission ($>1.8$~ns after the pump pulse) corresponds to photons scattered by the atoms. A coincidences-to-accidentals ratio (CAR) of 20 is observed when correlating signal-idler photon statistics. The CAR is reduced to 4.2 for the atoms-idler correlations, but still clearly shows that signal-idler photon statistics are successfully imparted upon the atomic ensemble. The raw correlations data corresponding to Fig.~\ref{fig_nlo}(f) is presented in Extended Data Fig.~\ref{fig_kerr}. We thus conclusively demonstrate the \textit{in situ} interaction of a parametric nonlinear process with a CQED system. 

\section*{Discussion}

In summary, we have developed an experimental platform for studying mesocopic CQED based on solid-state defects in WGM resonators. Due to the unique combination of the long photon lifetime in the cavity and the narrow inhomogeneous distribution of the atoms, we reach the single-emitter strong coupling regime and observe collective interference of $\sim10$ emitters. This mesoscopic regime is evidenced by the observation of simultaneously bunched ($g^{(2)} > 1$) and non-classical ($g^{(2)} < 1$) statistics from the few-atom ensemble. Realizing for the first time a two-mode optical system with spectral and spatial disorder within an atomic ensemble, we observe and theoretically study the steady-state symmetry breaking (chirality) emergent in disordered CQED systems. Looking ahead, the powerful technique of the recently-demonstrated check-probe spectroscopy\cite{van2024check} can enable the direct observation of strongly-coupled emitter dynamics, overcoming the fast spectral diffusion of atoms in the present experiment. Ultimately, through spatial control of emitter placement in the resonator\cite{day2023laser} and electric control of the emitter detunings \cite{lukin2020spectrally}, we can transition from the study of static but uncontrolled disorder to programmable disordered systems.

In demonstrating CQED within a nonlinear resonator via an in-situ interaction of a Kerr-nonlinear process with an atomic ensemble, we provide an experimental platform which can be used for the realization of theoretical protocols \cite{leroux2018enhancing, le2024cavity, qin2024quantum, lau2025efficient} and the observation of physical phenomena \cite{gardiner1986inhibition, carmichael1987resonance} related to nonlinear pair generation and squeezed light in CQED. The system may be engineered to generate atom-atom entanglement directly via degenerate pair generation (\textit{i.e.}, bichromatic pumping at modes $-1$ and $+1$ to generate a pair at mode 0). Exploring the second-order nonlinearity (present in noncentrosymmetric crystals such as SiC) for the parametric drive term may also be worthwhile due to its much larger intrinsic strength.

\textbf{Acknowledgments}
We thank Eric I. Rosenthal, Aashish Clerk, Tian Zhong, Joonhee Choi, Rahul Trivedi, Michael J. Feldman, Ignacio Cirac, Dominik Wild and Kiyoul Yang for helpful discussions. We thank NGK Insulators for SiCOI material for photonic probe fabrication. This work has been supported by the Vannevar Bush Faculty Fellowship from the US Department of Defense. DML acknowledges support of the Harvard Quantum Initiative Postdoctoral Fellowship. BW and MB acknowledge support from from the Munich Center for Quantum Science and Technology (MCQST), funded by the Deutsche Forschungsgemeinschaft (DFG) under Germany’s Excellence Strategy (EXC2111-390814868). JUH acknowledges support from Swedish Research Council (VR 2020-05444) and European Union’s HORIZON project SPINUS (101135699). TO acknowledges support from SIP 3rd, Promoting Application of Advanced Quantum Technologies to Social Challenges. GS acknowledges support from the Stanford Bloch Postdoctoral Fellowship. TKL acknowledges support from the NSF Graduate Research Fellowship. Part of this work
was performed at the Stanford Nano Shared Facilities
(SNSF)/Stanford Nanofabrication Facility (SNF), sup-
ported by the National Science Foundation under award
ECCS-2026822.

\section*{Methods}

\renewcommand{\figurename}{Extended Data Fig.}
\setcounter{figure}{0}
\textbf{Fabrication of WGM resonators.} The 4H-SiCOI material stack is prepared as in Ref.~\cite{lukin2023two}, with the same bulk SiC starting material (a 20~\textmu m n-doped epilayer with a nitrogen concentration $2\cdot 10^{13}$~cm\textsuperscript{-3}), with the difference of 35 times greater electron irradiation dose (fluence $3.5\cdot10^{14}$~cm$^{-2}$ at 2~MeV), to generate a larger density of V\textsubscript{Si} defects. The devices are fabricated in SiCOI via photolithographic pattern transfer akin to the process first reported in Ref.~\cite{jin2022micro}, but utilizing a short (15~s) bake at 135\textdegree C in ambient atmosphere to achieve structures with a small radius of curvature. The surface roughness of the resonators is measured via atomic force microscopy. After pattern transfer, the resonator is undercut via a vapor hydroflouric acid etch of the SiO\textsubscript{2} and a gaseous xenon difluoride etch of the Si.

\textbf{Fiber-interfaced SiC waveguide probe.} The detailed presentation of the development of the fiber-interfaced waveguide probe \cite{catanzaro2023cryogenic} are presented in separate publication (in preparation). For this experiment, the waveguide dimension is optimized for single-mode operation in TE and TM polarization. In the experiments presented in this work, a probe with a total transmission efficiency of 15\% (corresponding to 39\% in- and out-coupling efficiencies) is used. To fabricate the probe we use high-purity semi-insulating 4H-SiCOI (NGK Insulators). Despite operating the probe in-contact with the resonator surface, due to the small size and mechanical compliance of the waveguide, no deterioration of the resonator performance over time has been observed.

\textbf{Extracting $g$ from strong coupling dynamics in presence of spectral diffusion.}
Above-resonant excitation of the emitter causes fluctuations of local charges within the silicon carbide crystal in emitter proximity. Changes in the local charge environment of the emitter alter its resonant frequency via dc Stark shift \cite{lukin2020spectrally}, causing spectral diffusion \cite{orphal2023optically}. This can be modeled statistically as sampling independently from a Gaussian distribution around the center frequency of the emitter. Since the V\textsubscript{Si} has two spin states manifolds with an optical transition splitting of 1~GHz \cite{liu2024silicon} due to the excited state zero-field splitting (ZFS), in absence of spin initialization, a bimodal distribution comprising of two Gaussian distributions separated by the excited state ZFS is used. In the case of single-photon transport, the emitter-cavity system is entirely linear and is modeled exactly via a system of coupled cavities. For each emitter detuning condition, the dynamics are solved numerically via an ordinary differential equation solver. To obtain the best fit for the emitter-cavity coupling $g$, two alternating gradient descent optimizations are used. The first optimization step optimizes the emitter coupling parameter and the emitter central frequency to match the photon occupation in the resonator cavity. The second step applies scaling factors to the simulated cavity occupation, including amplitude scaling, dark-counts noise offset, and excitation time offset.  Both optimizations are done using gradient descent with the L-BFGS-B method.

\begin{figure*}[t]
\includegraphics[width=0.9\textwidth]{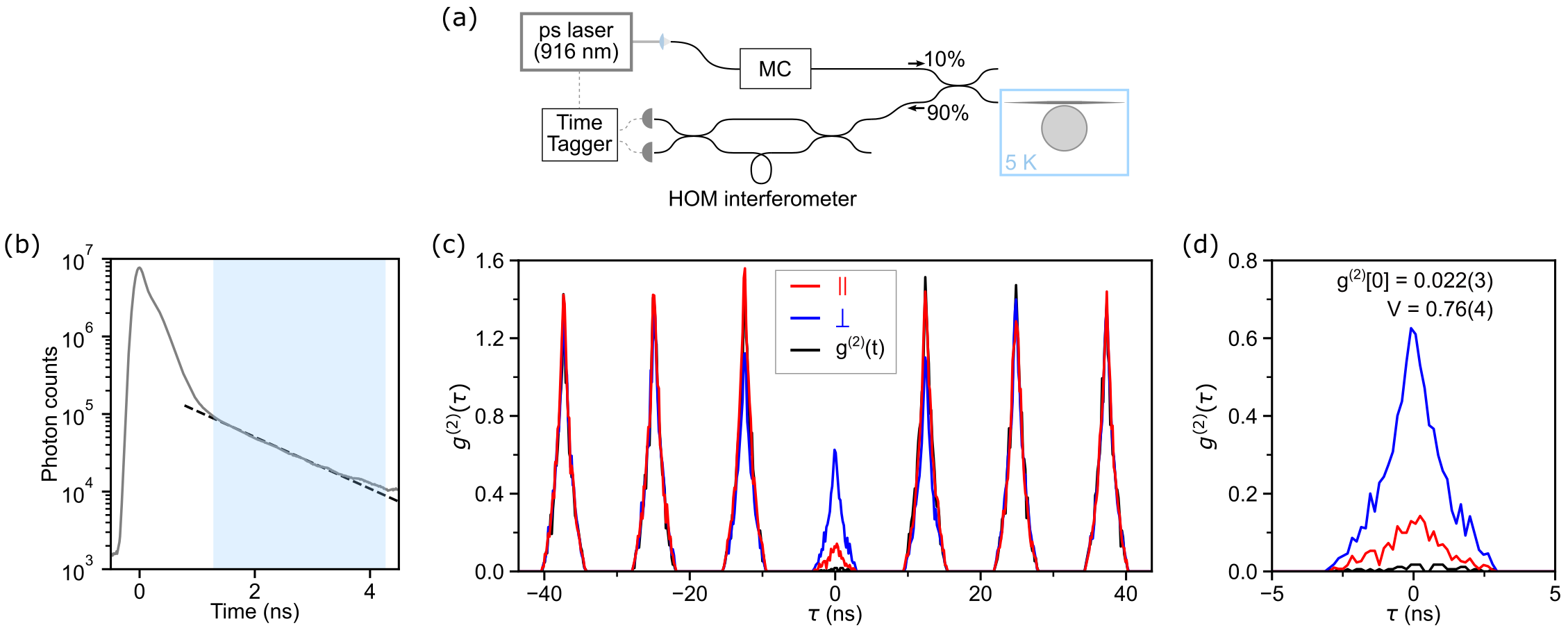}
\centering
\caption{
\textbf{Hong-Ou-Mandel (HOM) measurement of consecutive photon indistinguishability.} \textbf{(a)} The outline of the experimental setup. A picosecond laser is pulse shaped with a monochromator (MC) and sent to the resonator coupled to a single emitter. The back-scattered signal is passed through a HOM unbalanced interferometer with the delay between the arms equal to the inter-pulse delay. Photons are detected on a pair of SNSPDs and correlated with a time tagger. \textbf{(b)} The temporal response of the system. The fast decay after the laser pulse arrival time corresponds to the direct back-scattering by the cavity. The atom emission used for indistinguishability measurment is indicated by the blue region. The dashed line is a fit to an exponential decay with a lifetime of 1.30~ns. \textbf{(c)} Photon correlation for parallel polarization (indistinguishable) configuration (red), perpendicular polarization (distinguishable) configuration (blue), and single photon purity configuration bypassing the HOM inteferometer (black). \textbf{(d)} Zoom-in of the center peak in (c). The visibility and the single photon purity are measured to be $V=0.76(4)$ and $g^{(2)}[0]=0.022(3)$.}
\label{fig_indistinguishability}
\end{figure*}

\textbf{Measurement of photon indistinguishability.} 
The experimental setup for the HOM measurement is presented in Fig.~\ref{fig_indistinguishability}(a). The cavity-coupled single emitter is excited coherently via a resonant picosecond laser shaped to a bandwidth of 3~GHz via a bandwidth-tunable double-pass monochromatic. The back-scattered emission is detected and passed through the HOM interferometer before detection on a pair of SNSPDs (PhotonSpot, Inc), correlated with a time tagger. No spectral filtering is performed on detection. The experiment is conducted in a configuration where the cavity is strongly over-coupled to the waveguide (loaded Q factor of $2.28\cdot10^5$), putting the CQED system into a weakly-coupled, Purcell-enhanced regime. The emitter lifetime is measured to be $1.3$~ns (Fig.~\ref{fig_indistinguishability}(b)), corresponding to a Purcell-enhanced decay rate $\Gamma$ of 122~MHz. We first measure the single-photon purity of the emission, and find minimal two-photon coincidence events, with $g^{(2)}[0]=0.022(3)$. We then measure the HOM interference in the distinguishable (perpendicular polarizations of the two interferometer paths) and indistinguishable (parallel polarizations) configurations. We find that the consecutively emitted photons are indistinguishable with a visibility value of $V=0.76(4)$. We note that the single-photon impurity contributions to reduction of $V$ is negligible. The absence of a ``dip'' near zero time-delay indicates lack of spectral diffusion on the inter-pulse delay timescales (12.5~ns) \cite{komza2022indistinguishable}. From the relation between $V$, $\Gamma$, and $\gamma'$,\cite{thoma2016exploring} $$V = \frac{\Gamma}{\gamma' + \Gamma},$$
we obtain $\gamma' \leq 39$~MHz. It is an upper bound on dephasing because of the assumption that all visibility degradation comes from dephasing, thus assuming perfect HOM balance and polarization alignment. From the intrinsic linewidths of the two-transitions in the V\textsubscript{Si} of 26~MHz and 14~MHz \cite{liu2024silicon}, we obtain an maximum total linewidth of 65~MHz.

\textbf{Emitter phase estimation.} The non-trivial relative phases $\{ \phi_i \}$ of emitters can be extracted using the fiber stretcher MZI shown in Fig.~\ref{fig_hamiltonian}(b), which stretches the fiber using an applied voltage $V$. To extract the phase from the MZI, a sinuisoidal fit model $A \sin(\Phi_i(V)) + B$ is used for both MZI traces, where $\Phi_i(V)$ is a polynomial function of $V$ and $A, B$ are the amplitude and the average of the MZI signal, respectively. Before fitting, a convolution is applied to smoothen the MZI traces. 
The fitting is performed first by extracting the peaks of the convolved MZI interference signals, and then $\Phi_i(V)$ is fitted against the peaks. The fitted parameters from $\Phi_i(V)$ is then used as initial guess for fitting $A \sin(\Phi_i(V)) + B$ against the full signal and reference MZI traces, which is performed through minimizing the mean square error function using Powell algorithm, an optimization algorithm that excels at optimizing noisy functions. The phase $\phi_i$ is extracted as the difference between the resulting fits of both traces. To estimate errors on the extracted phases, we initialize SciPy's curvefit with the fitted parameters and then obtain the diagonal of the fit's covariance matrix, thereby finding the error bar for phases. 

\begin{figure*}[t]
\includegraphics[width=0.65\textwidth]{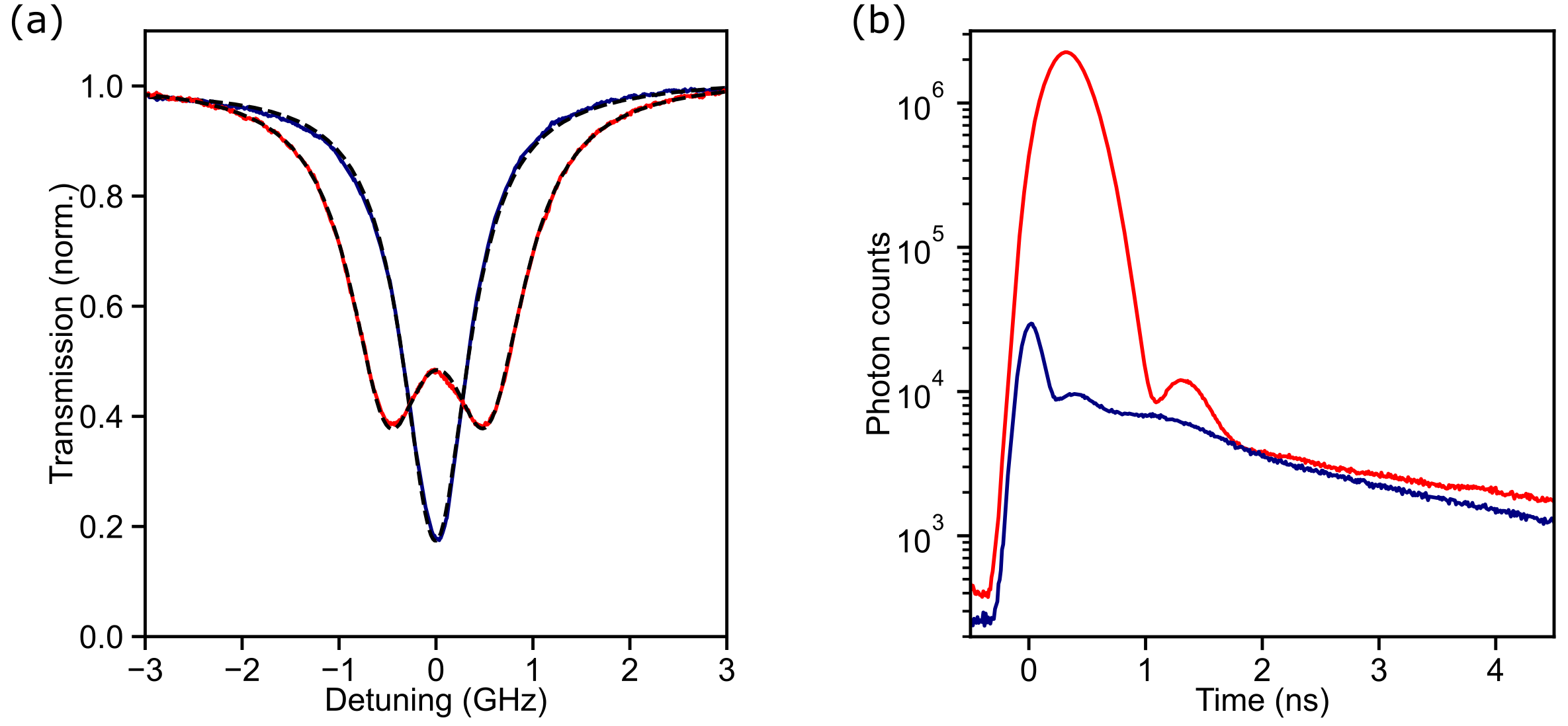}
\centering
\caption{
\textbf{Effect of large $g_{\text{bs}}$ on back-scattering dynamics.} \textbf{(a)} Transmission spectra for the same cavity mode with different strengths of $g_{\text{bs}}$ controlled by the waveguide coupling condition. The detuning is relative to $327.10597$~THz. Shown in blue is the condition where modes $\hat a$ and $\hat b$ are coupled weakly ($g_{\text{bs}}/\kappa = 0.053$). Shown in red is the case where the coupling is an order of magnitude stronger, $g_{\text{bs}}/\kappa = 0.54$. \textbf{(b)} The back-scattering photo dynamics corresponding to the two cases.} 
\label{fig_g_bs}
\end{figure*}

\textbf{Effect of large $g_{\text{bs}}$ on back-scattering dynamics.} The ability to carefully control the coupling conditions of the waveguide to the cavity via the fiber-interfaced waveguide probe makes it possible to not only control $\kappa_C$ but also $g_{\text{bs}}$, since the waveguide can break the resonator symmetry significantly. Supplementing the experiment shown in Fig.~\ref{fig_hamiltonian}(e), which illustrated the effect of $g_{\text{bs}}$ in the regime where it is commensurate with the effective back-scattering rate by an atom, we demonstrate what happens when $g_{\text{bs}}$ dominates. As shown in Fig.~\ref{fig_g_bs}, we prepare the cavity resonance in two different conditions, and compare the back-scattering dynamics for the two cases. For the case where $g_{\text{bs}}$ is strong, the direct cavity back-scattering overwhelms the atom-induced backscattering, as expected. Observation of quantum temporal dynamics is thus clearly more favorable in the regime of weak $g_{\text{bs}}$.

\textbf{Bad-cavity master equation.} 
The effective master equation presented in the main text is based on the method proposed in Ref.~\onlinecite{jaeger2022lindblad}, which we extend to include the intrinsic emitter dissipation. 
Specifically, we assume that the state of the whole system, $\hat \rho$, evolves according to the master equation
\begin{equation}
    \partial_t \hat\rho = -i[\hat H_e + \hat H_{\rm cav} + \epsilon \hat H_{\rm int}, \hat\rho] + \epsilon^d \mathcal{D}_e \hat\rho + \mathcal{D}_{\rm cav} \hat\rho \,, \label{eq:generalME}
\end{equation}
where
\begin{equation*}
    \hat H_{\rm cav} = \sum_{k=1}^K \omega_k \hat a^\dag_k \hat a_k\,, \quad \mathcal{D}_{\rm cav} = \sum_{k=1}^K \kappa_k \mathcal{D}[\hat a_k] \,,
\end{equation*} 
and the emitter-cavity interaction is of the form
\begin{equation*}
    \hat H_{\rm int} = \sum_{k=1}^K \hat a^\dag_k \hat S_k + {\rm H.c.}
\end{equation*}
Here, $\hat a_k$ denotes the annihilation operator of the $k$-th bosonic mode, while $\hat S_k$ is an operator of one or several emitters, which we leave unspecified for the moment. 
We also leave $\hat H_e$ and $\mathcal{D}_e$  unspecified for the moment. 
The small, dimensionless parameter $\epsilon$ makes explicit the separation between the timescales of bosonic decay, emitter-cavity interaction, and emitter decay. 
We discuss the cases $d = 1$ and $d = 2$ in parallel, since both lead to essentially the same effective master equation, although there are subtle differences between the two cases when computing the photon statistics.

In the limit $\epsilon\to 0$, the emitter and cavity subsystems evolve independently. 
Any product state of the form $\hat\rho(t_0) = \hat\rho_e(t_0) \otimes \ket{\vac}\!\bra{\vac}$, where $\ket{\vac}$ denotes the vacuum state of all the bosonic modes, and $\hat\rho_e$ denotes a density operator of the emitter subsystem, will remain of the same form at any later times $t>t_0$. 
In fact, any steady state is of this form.
For $\epsilon > 0$, this is no longer the case, since $\epsilon \hat H_{\rm int}$ couples the vacuum state to higher photon-number states. Nonetheless, we can find a unitary transformation
\begin{equation*}
    \hat D = \exp\left(\epsilon\sum_{k} \hat a^\dag_k \hat\alpha_k - {\rm H.c.}\right) \,,
\end{equation*}
where $\hat\alpha_k$ is an emitter operator satisfying
\begin{equation}
    i[\hat H_S, \hat\alpha_k] + i\omega_k \hat\alpha_k + i \hat S_k + \frac{\kappa_k}{2} \hat\alpha_k = 0 \,, \label{eq:condition_alpha}
\end{equation}
such that in the transformed frame the dynamics of any state within the zero-photon subspace decouples---up to terms that are $\bigO(\epsilon^3)$---from any state with higher photon number.
Concretely, by expanding the system's Liouvillian in the transformed frame (denoted here by a tilde) as a power series in $\epsilon$, one can show that a state
\begin{multline}
    \tilde\rho(t_0) = \hat\rho_e(t_0)\otimes\ket{\vac}\!\bra{\vac} \\ 
    + \epsilon^2 \sum_{\bm{n}\neq\vac}\left(\hat\rho_{e,\bm{n}}(t_0) \otimes\ket{\bm{n}}\!\bra{\vac} + {\rm H.c.}\right) + \bigO(\epsilon^3) \,, \label{eq:form_rho_tilde}
\end{multline}
will retain the same form at any later times, $t > t_0$.
Furthermore, the dynamics of $\hat\rho_e(t)$ up to terms $\bigO(\epsilon^3)$ is given by $\partial_t \hat\rho_e = \mathcal{L}_{\rm eff} \hat\rho_e$, with
\begin{gather*}
    \mathcal{L}_{\rm eff} \hat\rho_e = -i[\hat H_{\rm eff}, \hat\rho_e] + \mathcal{D}_{\rm eff} \hat\rho_e \,, \\
    \hat H_{\rm eff} = \hat H_e + \frac{\epsilon^2}{2} \sum_{k} \left(\hat\alpha^\dag_k \hat S_k + \mathrm{H.c.}\right) \,, \\
    \mathcal{D}_{\rm eff} = \epsilon^d \mathcal{D}_e  + \epsilon^2\sum_k \kappa_k \mathcal{D}[\hat\alpha_k] \,.
\end{gather*}
For a state of the general form above, $\tr_{\rm cav} \tilde\rho = \hat\rho_e + \bigO(\epsilon^3)$.

\begin{figure*}[t]
\includegraphics[width=0.8\textwidth]{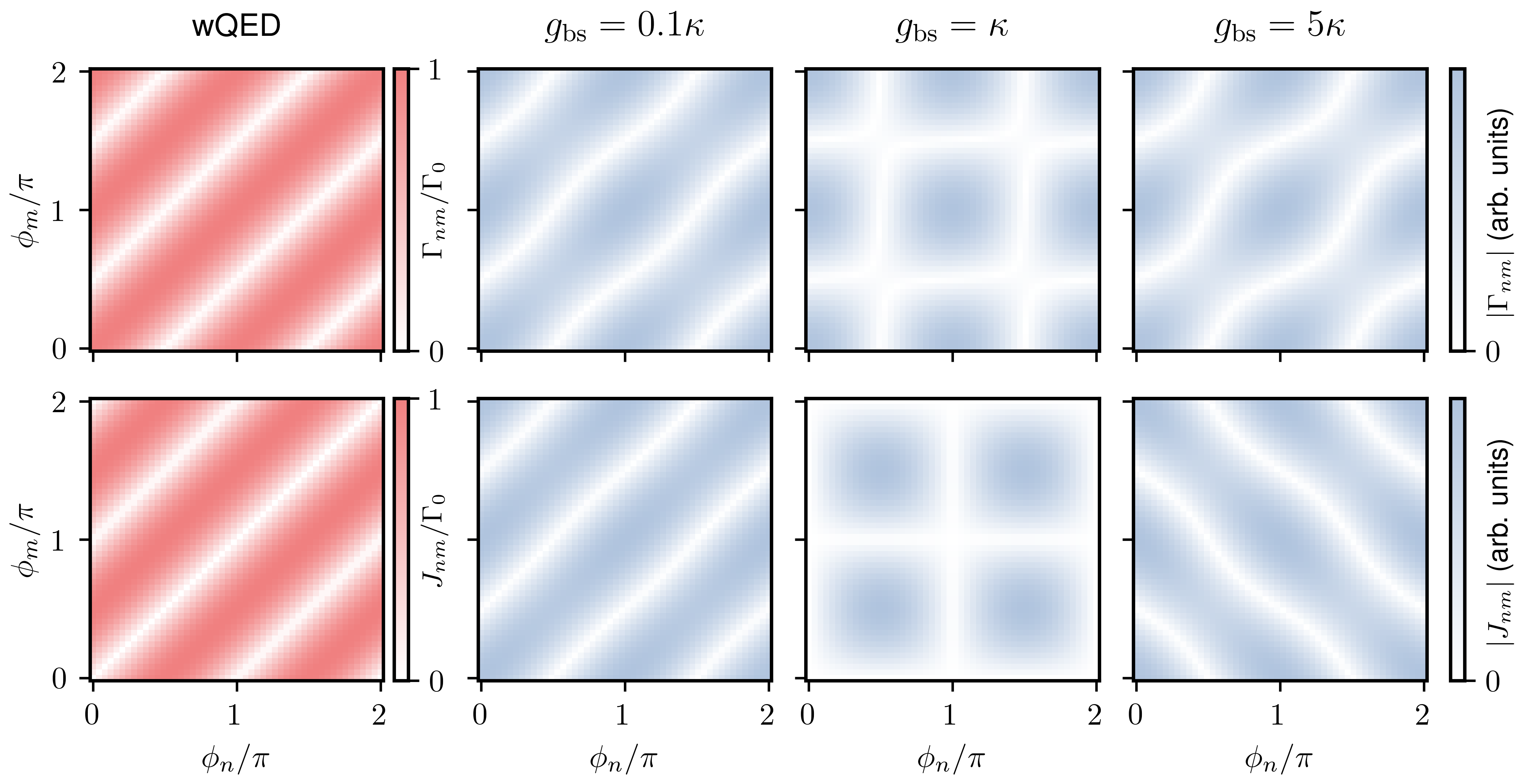}
\centering
\caption{\textbf{Markovian interaction strengths.} Comparison of the Markovian coherent and dissipative interaction strengths in wQED, with $\Gamma_0$ denoting the emision rate into the waveguide~\cite{chang_cavity_2012,pichler_quantum_2015}, and for our bad-cavity model with varied direct coupling strengths $g_{\rm bs}\neq 0$. Note that we plot only the absolute value of the interaction strengths, however the interactions are also associated with a non-trivial complex phase which cannot be removed by a gauge transformation (see Methods). In all panels, we assume identical emitters with $\Delta_n=\kappa$ and $g_n=0.01\kappa$.}
\label{fig_interactions}
\end{figure*}

We now adapt these equations to the case where $\hat H_e = \sum_n \Delta_n \hat\sigma^\dag_n \hat\sigma_n$, and $\epsilon S_k = \sum_n g_{nk} \hat\sigma_n$. 
Indeed, the original master equation in the main text can be written in this way if it is expressed in terms of the cavity eigenmodes $\hat a_1 = (\hat a + \hat b)/\sqrt{2}$ and $\hat a_2 = (\hat a - \hat b)/\sqrt{2}$. In a frame rotating with the average mode frequency $\omega_{\rm cav}$, we have 
$\Delta_n = \omega_n^e - \omega_{\rm cav}$, $\omega_1 = g_{\rm bs}$, and $\omega_2 = -g_{\rm bs}$. The emitter-mode couplings and mode decay rates are $g_{n1} = g_n\cos(\phi_n)$, $g_{n2} = -ig_n\sin(\phi_n)$, and $\kappa_1 = \kappa_2 = \kappa$.
A solution of Eq.~\eqref{eq:condition_alpha} is given in this case by 
\begin{equation}
    \epsilon \hat\alpha_k = \sum_n c_{nk} \hat\sigma_n \quad\text{with}\quad c_{nk} = \frac{g_{nk}}{\Delta_n - \omega_k + i\kappa_k/2} \,. \label{eq:alphak}
\end{equation}
The effective emitter Hamiltonian $\hat H_{\rm eff}$ and dissipator $\mathcal{D}_{\rm eff}$ can then be written as shown in the main text, with
\begin{equation*}
    J_{mn} = \sum_k \frac{c^*_{mk} g_{nk} + c_{nk} g^*_{mk}}{2} \,, \quad \Gamma_{mn} = \sum_k \kappa_k c_{mk} c^*_{nk} \,.
\end{equation*}
For $g_{\rm bs} = 0$, these coupling matrices can be made real through a gauge transformation: $\hat\sigma_n \to e^{i\theta_n} \hat\sigma_n$ with $\theta_n = \arg(\Delta_n + i\kappa/2)$. 
For $g_{\rm bs}\neq 0$, this is not generally possible, as can be seen from the values of gauge-invariant quantities, such as the products $J_{lm} J_{mn} J_{nl}$ for pairwise distinct indices $(l, m, n)$, which are complex in general. In Extended Data Fig.~\ref{fig_interactions}, we show illustrative examples of the non-trivial effective interaction patterns that can be realized for $g_{\rm bs}\neq 0$.

\begin{figure*}[t]
    \centering
    \includegraphics{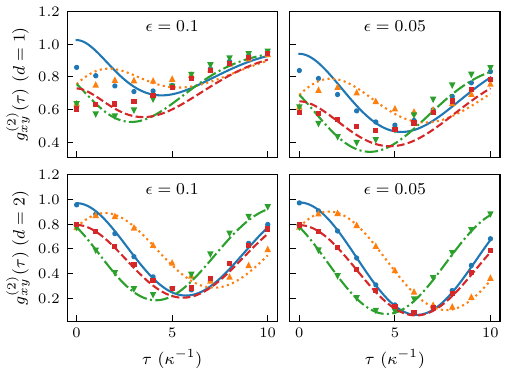}
    \caption{\textbf{Second-order coherence functions for two-emitter system.} Steady-state photon correlations for two emitters with system parameters $\Delta_1 = \kappa$, $\Delta_2=\kappa/2$, $\phi_1 = \pi/4$, $\phi_2=0$, $g_{\rm bs}=\kappa/2$, $g_n=\epsilon\kappa$, and $\gamma_n=\gamma'_n=\gamma^{\rm ex}_n=\epsilon^d\kappa$ for $n=1,2$.
    Lines correspond to the numerically exact result computed with the full model, while markers correspond to the approximation based on the effective emitter-only master equation.}
    \label{fig:g2_comparison}
\end{figure*}

\textbf{Average photon number and photon statistics}
From $\hat\rho_e$ alone we can compute the mean photon number in each cavity mode. 
Since
\begin{multline*}
    \tilde a_k = \hat a_k + \epsilon \hat\alpha_k \\
    + \frac{\epsilon^2}{2} \sum_l \left([\hat\alpha^\dag_l, \hat\alpha_k]\hat a_l + \hat a^\dag_l [\hat\alpha_k,\hat\alpha_l]\right) + \bigO(\epsilon^3) \,,
\end{multline*}
for $\tilde\rho$ of the form shown in Eq.~\eqref{eq:form_rho_tilde}, we have
\begin{equation*}
    \tilde a_k \tilde\rho \tilde a^\dag_k = \epsilon^2 \hat\alpha_k \hat\rho_e \hat\alpha^\dag_k \otimes \ket{\vac}\!\bra{\vac} + \bigO(\epsilon^3) \,.
\end{equation*}
Thus, to leading order in $\epsilon$, $\mean{\hat a^\dag_k \hat a_k} \approx \epsilon^2 \tr(\hat\alpha^\dag_k \hat\alpha_k \hat\rho_e)$. By the Quantum Regression Theorem, the second-order photon correlations can be computed as
\begin{equation*}
    \mean{\hat a^\dag_k \hat a^\dag_j(\tau) \hat a_j(\tau) \hat a_k} = \tr\left(\hat a_j e^{\mathcal{L}\tau}\big(\hat a_k \hat\rho \hat a_k^\dag\big)\hat a^\dag_j\right) \,.
\end{equation*}
Evolving the steady state after the detection of a photon in the $k$-th mode, we obtain
\begin{equation}
    e^{\tilde{\mathcal{L}}\tau}(\tilde a_k \tilde\rho \tilde a_k^\dag) = \epsilon^2 e^{\mathcal{L}_{\rm eff}\tau}(\tilde\alpha_k \tilde\rho_e \tilde\alpha_k^\dag)\otimes \ket{\vac}\!\bra{\vac} + \bigO(\epsilon^3) \,. \label{eq:evolved_jump_ss}
\end{equation}
From the first term on the right hand side, the leading contribution to the second-order photon correlations is
\begin{equation*}
    \epsilon^4 \tr\left(\hat\alpha_j e^{\mathcal{L}_{\rm eff}\tau} \big(\hat\alpha_k \hat\rho_e \hat\alpha_k^\dag\big)\hat\alpha^\dag_j\right) \,,
\end{equation*}
however, there may be contributions of lower or equal order stemming from the remaining terms.
Only when terms $\propto \ket{\bm{n}}\!\bra{\bm{n}}$ for $\bm{n}\neq\vac$, and terms $\propto \hat a^\dag_j \ket{\bm{n}}\!\bra{\bm{n}}, \ket{\bm{n}}\!\bra{\bm{n}}\hat a_j$ on the right-hand side of Eq.~\eqref{eq:evolved_jump_ss} are at least $\bigO(\epsilon^5)$ and $\bigO(\epsilon^4)$, respectively, we can approximate, 
to leading order in $\epsilon$,
\begin{equation}
    \mean{\hat a^\dag_k \hat a^\dag_j(\tau) \hat a_j(\tau) \hat a_k} \approx
    \epsilon^4 \tr\left(\hat\alpha_j e^{\mathcal{L}_{\rm eff}\tau} \big(\hat\alpha_k \hat\rho_e \hat\alpha_k^\dag\big)\hat\alpha^\dag_j\right) \,. \label{eq:2nd-order_corr_approx}
\end{equation}
This is likely not the case for the particular model we are considering with $d=1$, since the steady state, of the form shown in Eq.~\eqref{eq:form_rho_tilde}, already contains terms $\propto \ket{\bm{n}}\!\bra{\bm{0}}$, with $\sum_k n_k = 1$, at order $\epsilon^2$.
For $d = 2$, by contrast, one can show, using the explicit expression~\eqref{eq:alphak} for $\hat\alpha_k$, that the steady state is actually of the form $\hat\rho_e\otimes\ket{\vac}\!\bra{\vac} + \bigO(\epsilon^3)$. 
To asses the validity of the approximation, in Extended Data Fig.~\ref{fig:g2_comparison} we compare the exact and the approximate values of the second-order coherence functions
\begin{equation}
\label{eq:g2_definition}
    g^{(2)}_{xy}(\tau) \approx \frac{\mean{\hat\alpha_x^\dag \hat\alpha_y^\dag(\tau) \hat\alpha_y(\tau) \hat\alpha_x}}{\mean{\hat\alpha^\dag_x \hat\alpha_x}\mean{\hat\alpha^\dag_y \hat\alpha_y}} \quad (x,y\in\{a,b\})\,,
\end{equation}
for a particular instance of the model, and different values of $\epsilon$ and $d$.
Note that the collective jump operators associated with the cw and ccw modes $a$ and $b$ are $\hat\alpha_a = (\hat\alpha_1 + \hat\alpha_2)/\sqrt{2}$ and $\hat\alpha_b = (\hat\alpha_1 - \hat\alpha_2)/\sqrt{2}$, respectively.

\textbf{Independent emitters.} Assuming that the approximation in Eq.~\eqref{eq:2nd-order_corr_approx} holds, the steady-state correlation functions can be expressed entirely in terms of emitter correlations through Eq.~\eqref{eq:g2_definition}. An analytical expression can be obtained under the assumption that the cavity-mediated collective interactions are negligible compared to the independent dynamics of each emitter. In this limit, the emitters are approximately uncorrelated in the steady state~\cite{grim2019scalable,auffèves_few_2011,kim_super-radiant_2018}, such that for the case $g_{\rm bs}=0$, the cross-correlations read
\begin{multline*}
    g^{(2)}_{\rm ab}(\tau) \approx 1 +\sum_n \frac{I_n^2}{I^2}\left(g_n^{(2)}(\tau)-1\right)
    \\+\sum_{n\neq m}\frac{I_nI_m}{I^2}\,\mathrm{Re}\left\{\,e^{i\,\mathrm{sign}(\tau)\phi_{nm}}\,g_n^{(1)}(\tau)g_m^{(1)}(-\tau)\right\}\,,
\end{multline*}    
with a corresponding expression for $g^{(2)}_{ba}$, and with the auto-correlations $g_{aa}(\tau)=g_{bb}(\tau)$ obtained by setting $\phi_n\to0$ in this expression. Here, we have defined $\phi_{nm}=2(\phi_n-\phi_m)$, as well as $I_n=\Gamma_n\langle\hat{\sigma}_n^\dagger\hat{\sigma}_n\rangle$ and $I=\sum_nI_n$, such that $I/I_n$ can be understood as providing a measure of the relative `participation' of the $n$-th emitter in the steady state. Additionally, $g^{(1)}_n(\tau)$ and $g^{(2)}_n(\tau)$ denote the normalised steady-state correlations of the $n$-th emitter, and since $g_n^{(1)}(0)=1$ and $g_n^{(2)}(0)=0$,
\begin{equation*}
    g^{(2)}_{\rm aa}(0) 
    =2-2\sum_nr_n^2\,.
\end{equation*}
Note that the term $\sum_nr_n^2$ is minimised for uniformly distributed values of $I_n$, and therefore $g^{(2)}_{\rm aa}(0)\leq 2(1-1/N)$, which reflects a quite general upper bound on the photon bunching from independent emitters~\cite{loudon_non-classical_1980}. The cross-correlations at $\tau=0$ can similarly be expressed as
\begin{equation*}
    g^{(2)}_{ab}(0)=g_{aa}^{(2)}(0) + \xi_\phi-1\,,
\end{equation*}
where $\xi_\phi=\sum_{nm}(I_nI_m/I^2)\cos(\phi_{nm})$ quantifies the phase disorder, as discussed in the main text. 

To construct the full time dependence of the correlations, we assume that the single-emitter correlations $g^{(1)}_n(\tau)$ and $g^{(2)}_n(\tau)$ can be approximated to lowest order by the well-known expressions for a two-level system subject to decay, dephasing and incoherent excitation~\cite{auffèves_few_2011,kim_super-radiant_2018}. In this case, the cross-correlations take the form
\begin{multline*}
    g^{(2)}_{\rm ab}(\tau) \approx 1 -\sum_n \frac{I_n^2}{I^2}\,e^{-(\gamma_n+\gamma_n^{\rm ex})\abs{\tau}}
    \\+\sum_{n\neq m}\frac{I_nI_m}{I^2} e^{-(\gamma_{nm}+\gamma_{nm}^{\rm ex}+\gamma_{nm}')\abs{\tau}}\cos\left(\Delta_{nm}\tau-\phi_{nm}\right)\,,
\end{multline*}
with $I_n=\gamma_n^{\rm ex}\Gamma_n/(\gamma_n+\gamma_n^{\rm ex})$. To lighten the notation, we have defined $\gamma_{nm}=(\gamma_n+\gamma_m)/2$, $\gamma_{nm}^{\rm ex}=(\gamma_n^{\rm ex}+\gamma_m^{\rm ex})/2$, $\gamma_{nm}'=(\gamma_n'+\gamma_m')/2$ and $\Delta_{nm}=\Delta_n-\Delta_m$. Once again, $g_{aa}(\tau)$ is obtained from this expression by setting $\phi_n\to0$, and we thus see that, at the level of independent emitters, the auto-correlations will always be symmetric with respect to $\tau=0$, while the cross-correlations may display some chirality in the presence of non-uniform detunings $\Delta_n$, as discussed in the main text.

If we assume identical emitters and neglect the dependence of the decay rates $\Gamma_n$ on the detunings $\Delta_n$, then $I_n=I/N$ and the effect of the spectral diffusion of the emitters can be easily incorporated by assuming that the frequencies $\Delta_n$ are sampled from independent normal distributions with mean values $\mu_n$ and standard deviations $s_n$. Averaging over these distributions, we obtain~\cite{grim2019scalable}
\begin{equation*}
\begin{split}
    &g_{ab}(\tau)=1-\frac{e^{-(\gamma+\gamma^{\rm ex})\abs{\tau}}}{N} \\
    &+\frac{e^{-(\gamma+\gamma^{\rm ex}+\gamma')\abs{\tau}}}{N^2}\sum_{n\neq m}e^{-s_{nm}^2\tau^2/2}\cos(\mu_{nm}\tau-\phi_{nm})\,,
\end{split}
\end{equation*}
where $\mu_{nm}=\mu_n-\mu_m$ and $s_{nm}=\sqrt{s_n^2+s_m^2}$. Of course, the assumption that $I_n=I/N$ will in general be violated due to the significant spread in emitter frequencies, which results in a non-negligible dependence of each $\Gamma_n$ (and hence $I_n$) on $\Delta_n$. Likewise, cavity Purcell enhancement of each emitter decay rate induces a further non-trivial frequency dependence in $g_n^{(1)}(\tau)$ and $g_n^{(2)}(\tau)$. With these corrections, the averaging over spectral diffusion can no longer be incorporated analytically, and the above equation should therefore be viewed pragmatically as a minimal model to study the effects of phase and spectral disorder on photon correlations from independent emitters, rather than a quantitatively accurate description of the experimental system.

\textbf{Numerical fit of $g^{(2)}_{aa}(\tau)$ and $g^{(2)}_{ab}(\tau)$.} In order to correctly model correlations even at the level of independent emitters, we must include in the model the metastable state of the V\textsubscript{Si} \cite{liu2024silicon}, which contributes a weak bunching away from $\tau = 0$ due to population shelving in the long lived dark state. Within the assumptions outlined in the previous section, this amounts to replacing the expressions for $g^{(1)}_n(\tau)$ and $g^{(2)}_n(\tau)$ with those for a three-level system. We denote the decay rates from the excited to the metastable state and from the metastable to the ground state of the $n$-th emitter by $\gamma^e_n$ and $\gamma^s_n$, respectively. To lighten the notation, we will also define $x_n=(\gamma_n+\gamma^e_n+\gamma^s_n+\gamma_n^{\rm ex})/2$, $y_n=(\gamma_n+\gamma^e_n)\gamma^s_n+(\gamma_n^s+\gamma_n^e)\gamma^{\rm ex}_n$, and $z_n=\sqrt{x_n^2-y_n}$. Before accounting for spectral diffusion, the cross-correlations then take the form
\begin{multline*}
    g^{(2)}_{\rm ab}(\tau) \approx 1 -\sum_n \frac{I_n^2}{I^2}\,e^{-x_n\abs{\tau}}\left(\cosh z_n\abs{\tau}-\lambda_n\sinh z_n\abs{\tau}\right)
    \\+\sum_{n\neq m}\frac{I_nI_m}{I^2} e^{-(\gamma_{nm}+\gamma_{nm}^e+\gamma_{nm}^{\rm ex}+\gamma_{nm}')\abs{\tau}}\cos\left(\Delta_{nm}\tau-\phi_{nm}\right)\,,
\end{multline*}
where we have further defined $\lambda_n=(x_n-\gamma_n^s+\gamma_n^e\gamma_n^{\rm ex}/\gamma_n^s)/z_n$ and $\gamma_{nm}^e=(\gamma_n^e+\gamma_m^e)/2$, and where now $I_n=\gamma_n^{\rm ex}\gamma_n^s\Gamma_n/y_n$. We will again assume identical emitters and neglect the dependence of the decay rates $\Gamma_n$ on the detunings $\Delta_n$. In addition, we will assume identical spectral diffusion across the emitters, so that we obtain
\begin{align*}
    g^{(2)}_{\rm ab}(\tau) \approx 1 
    &-\frac{1}{N}e^{-x\abs{\tau}}\left(\cosh z\abs{\tau}-\lambda\sinh z\abs{\tau}\right)
    \\
    &+\left(\xi_\phi-\frac{1}{N}\right)\,e^{-(\gamma+\gamma^e+\gamma^{\rm ex}+\gamma')\abs{\tau}}\,e^{-s^2\tau^2/2}\,.
\end{align*}
The auto- and cross-correlation data in Fig.~\ref{fig_single_emitter}(b) are fit using this simplified expression with free parameters $\{\gamma, \gamma', \gamma^{\text{ex}}, \gamma^e, N, \xi_\phi \}$. The spectral diffusion parameter $s$ is fixed at 1~GHz (corresponding to the cavity linewidth). The fit to the data includes a convolution with the detector timing jitter (82~ps).

\textbf{Steady-state chirality toy models.} As explained in the main text, for a system as complex as the one at hand, one generally would not expect 
$g^{(2)}_{aa}(\tau) = g^{(2)}_{bb}(\tau)$, and $g^{(2)}_{ab}(\tau) = g^{(2)}_{ba}(\tau) = g^{(2)}_{ab}(-\tau)$,
except in certain finely-tuned cases, such as when the steady state of the system is unique and the system has a weak symmetry \cite{Buca2012} that exchanges the two resonator modes.
Determining whether such a symmetry exist, or more generally, whether the correlation functions display the aforementioned behavior, is not as straightforward as one might assume.
Here, we discuss a few illustrative examples.
For the sake of simplicity, we assume that the dissipative part of the master equation is permutationally invariant,
i.e., all emitter decay rates, pumping rates, and dephasing rates are equal.
Consequently, we limit our analysis to the effects of disorder in the system Hamiltonian.
Furthermore, we also assume $g_{\rm gs} = 0$, since otherwise the steady will be chiral as long as there is at least one complex phase different from 0 (mod $\pi$).

\emph{Two-emitter system}---In the case of two emitters, a gauge transformation lets us split the complex phases evenly among all the couplings.
We can express
\begin{equation}
    \hat H = \sum_{n=1,2} \frac{\Delta_n}{2} \hat \sigma^\dag_n \hat \sigma_n +  \hat S_\phi^\dag \hat a + \hat S_{-\phi}^\dag \hat b + \mathrm{H.c.} \,,
\end{equation}
with $\hat S^\dag_\phi = (g_1 e^{i\phi}\hat\sigma^\dagger_1 + g_2 e^{-i\phi}\hat\sigma^\dagger_2)/\sqrt{2}$, $\phi = (\phi_1 - \phi_2)/2$.
Exchanging $\hat a \leftrightarrow \hat b$ is equivalent to exchanging $\phi \leftrightarrow - \phi$. 
Now, if $\Delta_1 = \Delta_2$ and $g_1 = g_2$, exchanging $\hat \sigma_1 \leftrightarrow \hat \sigma_2$ is also equivalent to changing $\phi \leftrightarrow - \phi$.
So, the system is invariant under the simultaneous exchange of $\hat a \leftrightarrow \hat b$ and $\hat \sigma_1 \leftrightarrow \hat \sigma_2$.
Since this is a symmetry transformation that permutes the two resonator modes, the second-order coherence functions are symmetric under the exchange of the two resonator modes.
On the other hand, if $\Delta_1 \neq \Delta_2$ and/or $g_1 \neq g_2$ this is no longer a symmetry transformation and the steady state is chiral.
Surprisingly, we find numerically that for $\Delta_1 = \Delta_2 = 0$ and $g_1 \neq g_2$, the second-order coherence functions are also symmetric.

\emph{Three-emitter system}---If $\Delta_1 = \Delta_2 = \Delta_3$ and $g_1 = g_2 = g_3$ it is possible that the system has a weak symmetry akin to the one described in the previous paragraph for specific values of the complex phases. 
For example, when one of the complex phases is half the sum of the other two (mod $2\pi$).
However, for arbitrary complex phases the steady state is typically chiral, even when the detunings and coupling strengths are the same for all the emitters. 
Again, if $\Delta_1 = \Delta_2 = \Delta_3 = 0$, we find numerically that the second-order coherence functions are symmetric, regardless the values of the remaining parameters.
We conjecture that this is always the case for any number of emitters, as long as all the detunings vanish.

\begin{figure*}[t]
    \centering
    \includegraphics{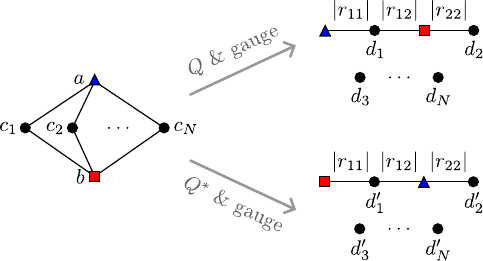}
    \caption{\textbf{Schematic illustration of homologous bosonic multi-emitter system.} On the left, graphical representation of the Hamiltonian in Eq.~\eqref{eq:bosonic_hamiltonian}. 
    The nodes correspond to different bosonic modes and the lines denote non-zero couplings.
    It is unitarily equivalent to the two systems on the right, in which the coupling strengths are the same (the absolute values of the entries of $R$ and $R^*$) but the $a$ and $b$ modes are interchanged.}
    \label{fig:schematics_bosonic}
\end{figure*}

\emph{Homologous bosonic system}---
If the steady state is such that the emitters remain mostly in their ground state, one may approximate them by additional bosonic modes $\hat\sigma_n \to \hat c_n$ for $n=1,\dots, N$. 
This results in a purely bosonic system that is much easier to simulate, requiring only a polynomial effort in the number of emitters \cite{Barthel2022}.
As we show in the following, for a system without dephasing there exist a different weak symmetry permuting the two resonator modes in the case where all detunings are equal (but not necessarily the coupling strengths).
Using matrix multiplication, we can express the Hamiltonian as
\begin{equation}
    \hat H = \frac{\Delta}{2} \sum_{n=1}^N \hat c^\dag_n \hat c_n + \begin{pmatrix}\hat c^\dag_1 & \cdots & \hat c^\dag_N\end{pmatrix} G
    \begin{pmatrix}
        \hat a \\ \hat b
    \end{pmatrix} + \mathrm{H.c.}\,, \label{eq:bosonic_hamiltonian}
\end{equation}
with
\begin{equation}
    G = \begin{pmatrix}
        g_1 e^{i\phi_1} \ g_1 e^{-i\phi_1} \\
        \vdots \\
        g_N e^{i\phi_N} \ g_N e^{-i\phi_N} \\
    \end{pmatrix} \,.
\end{equation}
We can define new orthonormal emitter modes such that mode $a$ only couples to one of them, say $d_1$, and mode $b$ only couples to two of them, $d_1$ and $d_2$, and they are decoupled from the rest of the emitter modes, $d_n$ for $n=3,\dots, N$.
This can be achieved with a QR decomposition, $G = QR$, where $Q$ is an $N\times N$ unitary and $R$ is an $N\times 2$ upper triangular matrix, defining
\begin{equation}
    \begin{pmatrix}\hat d^\dag_1 & \cdots & \hat d^\dag_N\end{pmatrix} = \begin{pmatrix}\hat c^\dag_1 & \cdots & \hat c^\dag_N\end{pmatrix} Q \,.
\end{equation}
Note that, since the pumping rates and decay rates are assumed to be the same for all the emitters, the dissipative part of the Liouvillian keeps the same form in the new basis.
Similarly, we can also define new orthonormal emitter modes such that $b$ is coupled to one of them, $a$ is coupled to two of them, and they are decoupled from the rest. 
The transformation and the new coupling strengths are given by the decomposition $G^* = Q^* R^*$.
Since $a$, $b$, and the two emitter modes they are coupled to are in a chain configuration (with open boundary conditions, see Fig.~\ref{fig:schematics_bosonic}), the complex phases of the couplings can be gauged out, so the two transformations lead to equivalent systems, with the $a$ and $b$ modes interchanged.
This implies the symmetry of the second-order coherence functions.
Note that this approach cannot be employed in the case of spins, since the linear combinations of spin operators defined would not correspond to new spin-1/2 degrees of freedom.

\bibliography{Reference}
\clearpage

\renewcommand{\thesection}{\Roman{section}}
\setcounter{section}{0}

\begin{figure*}[t]
\includegraphics[width=1\textwidth]{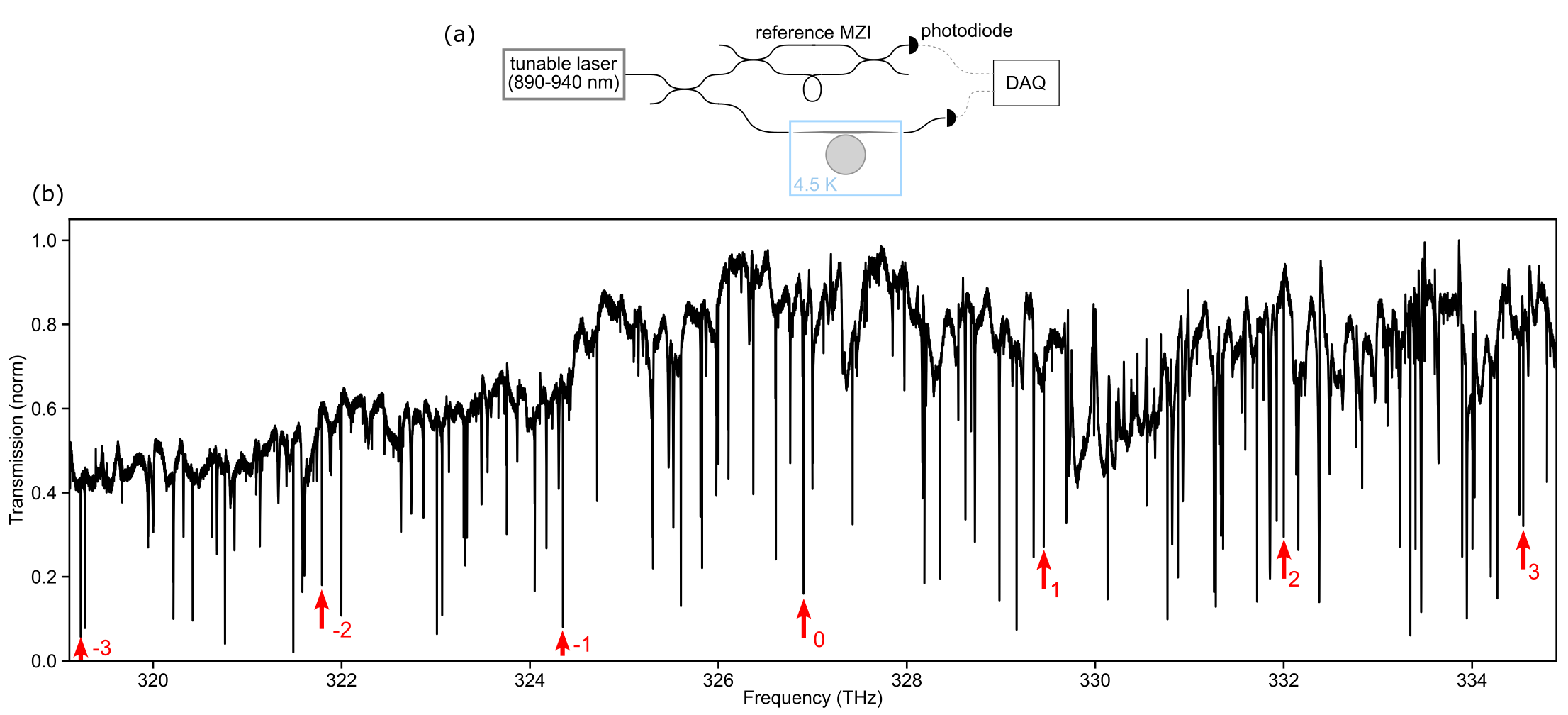}
\centering
\caption{\textbf{Transmission spectroscopy of the WGM resonator.} (a) Precisely-calibrated laser transmission scan of the resonator is performed via a scanning a mode-hop-free external-cavity diode laser while simultaneously acquiring time-domain transmission through the device and a Mach-Zehnder fiber interferometer (MZI). The transmission of the MZI serves as a calibrated ``ruler'' in the frequency domain allowing to map the time domain signal into the spectral domain. (b) Transmission scan across the tunable laser scanning range. The TM mode family and the azimuthal mode numbers (relative to the atom-coupled mode) is indicated with red arrows.}
\label{fig_mzi}
\end{figure*}

\begin{figure*}[t]
\includegraphics[width=1\textwidth]{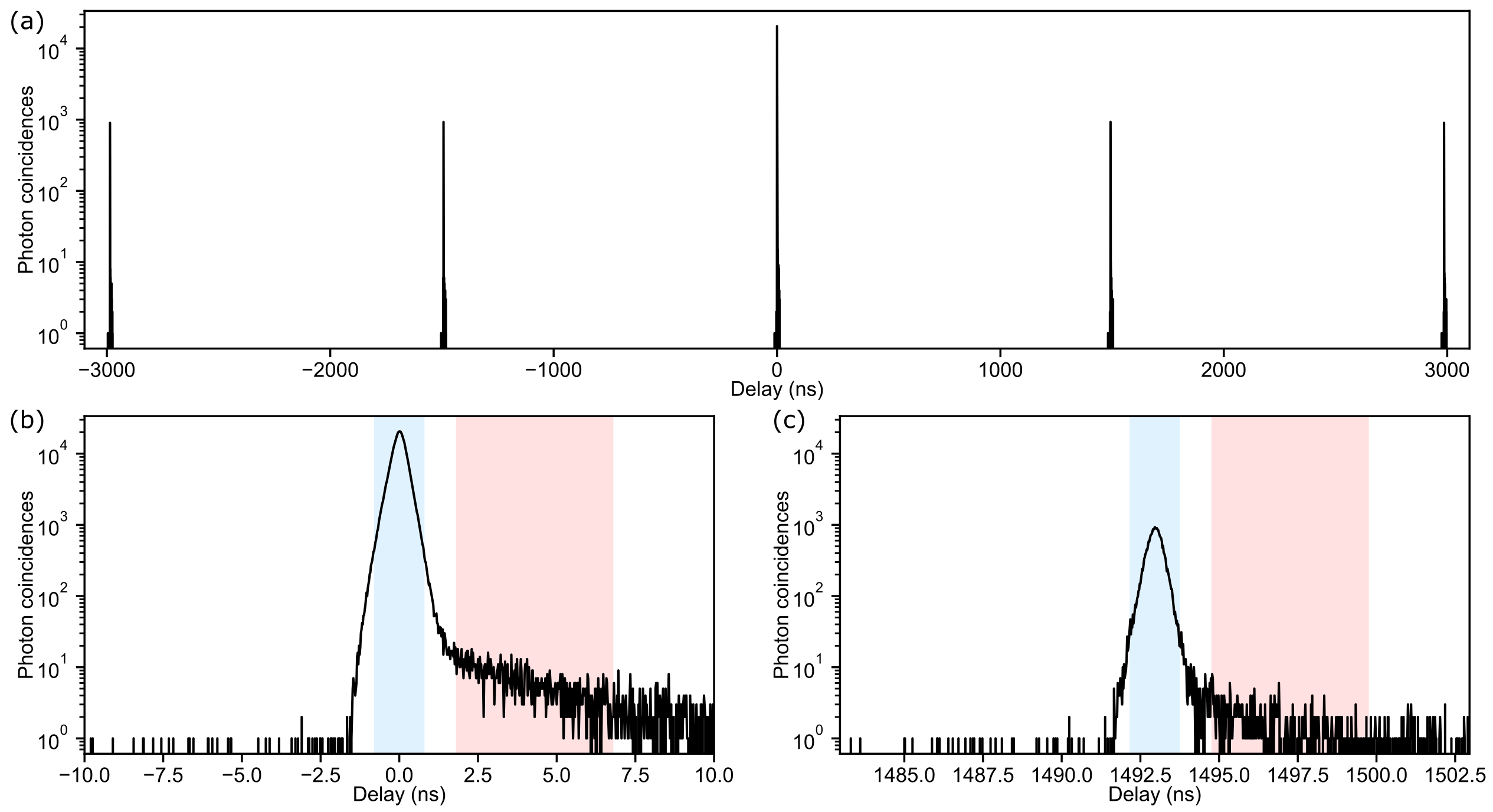}
\centering
\caption{\textbf{Raw two-photon coincidence data of the parametrically-driven atom ensemble.} \textbf{(a)} Correlations between emission into the signal and idler cavity modes shown for delay spanning $\pm2$ periods of the pump laser. \textbf{(b,c)} A zoom-in view of photon events the zero- and one-period delay, respectively. The integration time windows for fast emission (cavity only) and slow emission (atom scattering) are indicated in blue and red, respectively.}
\label{fig_kerr}
\end{figure*}

\end{document}